  \documentclass[journal,12pt,onecolumn,draftclsnofoot,]{IEEEtran}
\usepackage{epsfig,rotating,setspace,latexsym,amsmath,epsf,amssymb,bm}
\usepackage{cite,graphicx,authblk,color}
\usepackage{xcolor}
\usepackage{hyperref}
\usepackage{caption}
 \usepackage{graphicx}
 \usepackage{subcaption}

\usepackage{float}

\usepackage{url}

\usepackage{mathtools,amssymb,lipsum}
\usepackage{amsmath,amssymb,amsfonts }
\usepackage{ dsfont }

\usepackage{ mathrsfs }

\newtheorem{theorem}{Theorem}
\newtheorem{lemma}{Lemma}
\newtheorem{corollary}{Corollary}

\usepackage{fancyhdr}
\title{Secure Retrospective Interference Alignment
\author{Mohamed Seif \hspace{10pt} Ravi Tandon  \hspace{10pt} Ming Li}
\affil{Department of Electrical and Computer Engineering\\
University of Arizona, Tucson, AZ, USA.\\
E-mail: {\{mseif, tandonr, lim\}}@email.arizona.edu}}

\begin{document}
\maketitle
\newcommand\blfootnote[1]{%
  \begingroup
  \renewcommand\thefootnote{}\footnote{#1}%
  \addtocounter{footnote}{-1}%
  \endgroup
}

\blfootnote{This work of was supported in part by  U.S. NSF through grants CCF-1559758 and CNS-1715947, CNS-1564477, and ONR YIP grant N00014-16-1-2650. Parts of this work have been presented  in  IEEE  ISIT 2018 \cite{seif2018}.}

\thispagestyle{empty}
\vspace{-1 cm}

\begin{abstract}
In this paper, the $K$-user interference channel with secrecy constraints is considered with delayed channel state information at transmitters (CSIT). We propose a novel { \textit{secure}} retrospective interference alignment scheme in which the transmitters carefully mix information symbols with artificial noises to ensure confidentiality.  Achieving positive secure degrees of freedom (SDoF) is challenging due to the delayed nature of CSIT, and the distributed nature of the transmitters. Our scheme works over two phases: phase one in which each transmitter sends information symbols mixed with artificial noises, and repeats such transmission over multiple rounds. In the next phase, each transmitter uses delayed CSIT of the previous phase and sends a function of the net interference and artificial noises (generated in previous phase), which is simultaneously useful for all receivers. These phases are designed to ensure the decodability of the desired messages while satisfying the secrecy constraints.  We present our achievable scheme for three models, namely: 1) $K$-user interference channel with confidential messages (IC-CM), and we show that  $\frac{1}{2} (\sqrt{K} -6) $ SDoF is achievable,  2) $K$-user interference channel  with an external eavesdropper (IC-EE), and 3) $K$-user IC with confidential messages and an external eavesdropper (IC-CM-EE). We show that for the $K$-user IC-EE, $\frac{1}{2} (\sqrt{K} -3) $ SDoF is achievable, and for the  $K$-user IC-CM-EE, $\frac{1}{2} (\sqrt{K} -6) $ is achievable. To the best of our knowledge, this is the first result on the $K$-user interference channel with secrecy constrained models and delayed CSIT that achieves a  SDoF which scales with {  $\sqrt{K}$,  square-root of  number of users.}
\end{abstract}

 \textbf{Index Terms:}  Interference channel, secure retrospective interference alignment,  secure degrees of freedom (SDoF), delayed CSIT.

\section{Introduction}
\label{sec:introduction}

Delayed CSIT can impact the spectral efficiency of  wireless networks, and this problem has received significant recent attention. Maddah Ali and Tse in \cite{maddah2012completely} studied the delayed CSIT model for the  $K$-user multiple-input single-output (MISO) broadcast channel,
and showed that the optimal sum degrees of freedom (DoF) is given by $K/(1 + \frac{1}{2} + \dots + \frac{1}{K} )$ which is strictly greater than one DoF (with no CSIT) and less than $K$ DoF (with perfect CSIT). For the $K$-user single-input single-output (SISO) X network, $\frac{K^{2}}{2K-1}$ is maximum DoF with perfect CSIT \cite{cadambe2008interference}. In \cite{ghasemi2011degrees}, Ghasemi et al. devised a transmission scheme for the X channel with
delayed CSIT, and  showed that for the
 $K$-user SISO X channel under delayed CSIT,  $\frac{4}{3} - \frac{2}{3(3K-1)}$ DoF are achievable.  The problem of  delayed CSIT for  interference channels has been studied in several works\cite{ghasemi2011degrees, maggi2012retrospective, abdoli2013degrees,  kang2013ergodic, maleki2011retrospective}. The main drawback of these schemes is that the achievable DoF \textit{does not} scale with the number of users. In a recent work \cite{castanheira2017retrospective}, a novel transmission scheme for the $K$-user SISO interference channel is presented  which achieves $\frac{\lfloor \sqrt{K }\rfloor}{2} $ DoF almost surely under delayed CSIT model. The result in \cite{castanheira2017retrospective} is particularly interesting, as it shows that the sum DoF for the $K$-user interference channel \textit{does} scale with $\sqrt{K}$, even with delayed CSIT.

Another important aspect in wireless networks is ensuring secure communication between transmitters and receivers.  Many seminal works in the literature (see comprehensive surveys \cite{yener2015wireless, sennurtandon2016, liang2009information}) studied the secure capacity regions for multi-user settings such as wiretap channel, broadcastl and interference channels.  Since the exact secure capacity regions for many multi-user networks are not known,  secure degrees of freedom (SDoF) for a variety of models have been studied in \cite{he2009k, koyluoglu2011interference, xie2010real, bassily2012ergodic, gou2008secure, nafea2013many, yang2015secure}. More specifically, for the $K$-user MISO broadcast channel with confidential messages, the authors in \cite{yang2015secure} showed that the optimal  sum SDoF with delayed CSIT is given by  $K/(1 + \frac{1}{2} + \dots + \frac{1}{K} + \frac{K-1}{K} )$. The achievability scheme is based on a modification of  the (insecure) Maddah Ali and Tse's scheme in \cite{maddah2012completely} along with a  key generation method which uses delayed CSIT. The expression of the sum SDoF in  \cite{yang2015secure} is almost the same as in \cite{maddah2012completely} except a penalty term due to confidentiality constraints. For the $K$-user SISO interference channel with confidential messages under perfect CSIT, Xie and Ulukus showed in \cite{xie2015secure} that the optimal sum SDoF is $\frac{K (K-1)}{2 K -1}$. Also, there are various other works for different CSIT assumptions such as  MIMO wiretap channel with no eavesdropper  CSIT \cite{mukherjee2017secrecy}, broadcast channel with alternating CSIT  \cite{mukherjee2017secure}.

In this work, we consider the $K$-user SISO interference channel with secrecy constraints and delayed CSIT.  More specifically, we study three channel models, namely: 1) $K$-user interference channel with confidential messages, 2)  $K$-user interference channel with an external eavesdropper, and 3) $K$-user interference channel with confidential messages and an external eavesdropper.
We focus on answering the following fundamental questions regarding these channel models: (a) is positive SDoF achievable for the interference channel with delayed CSIT?, and (b) if yes, then how does the SDoF scale with $K$, the number of users?

\textbf{Contributions:}  We answer the above  questions for all the three channel models in the affirmative by showing that positive SDoF is indeed achievable for all these models. We show that for the $K$-user interference channel with confidential messages (IC-CM), $\frac{1}{2} (\sqrt{K} - 6)$ SDoF is achievable. Also, we show that for the $K$-user interference channel with an external eavesdropper (IC-EE), $\frac{1}{2} (\sqrt{K} -3) $ SDoF is achievable, and for the  $K$-user with confidential messages and an external eavesdropper (IC-CM-EE), $\frac{1}{2} (\sqrt{K} -6) $ is achievable. In Table $1$, we summarize the main results for the $K$-user IC under various secrecy constraints, and different three CSIT assumptions (i.e., perfect CSIT, delayed CSIT and no CSIT).

\begin{figure}[!h]
\centering
\includegraphics[width=3.6in, height=2.5in]{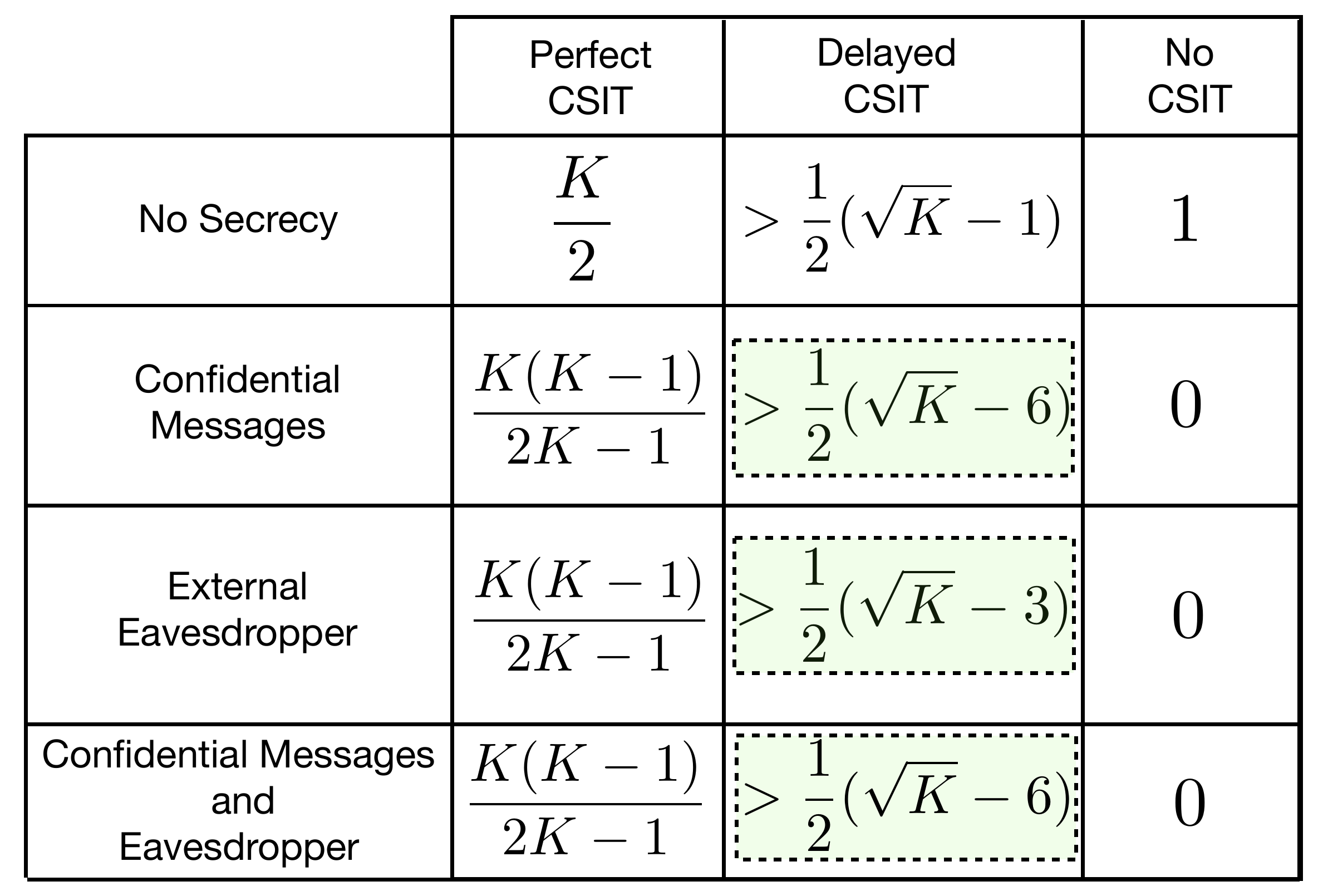}
  \caption*{Table 1: Summary of results on the $K$-user interference channel with different secrecy constraints and CSIT models. The highlighted results are from this paper. These results show that sum SDoF  scales with {$\sqrt{K}$. }}
\label{motivation}
\end{figure}

These results highlight the fact that in presence of delayed CSIT, there is almost no DoF scaling loss due to the secrecy constraints in the network compared to the no secrecy case \cite{castanheira2017retrospective}.  Our main contribution is a novel { \textit{secure}} retrospective interference alignment scheme, that is specialized for the interference channel with delayed CSIT. Our transmission scheme is inspired by the work of \cite{castanheira2017retrospective} in terms of the organization of the transmission phases. One of the main differences is  that the transmitters mix their information symbols with artificial noises so that the signals at each unintended receiver are completely immersed in the space spanned by artificial noise. However, this mixing must be done with only delayed CSIT,  and it should also allow successful decoding at the respective receiver.  Our scheme works over two phases: phase one in which each transmitter sends information symbols mixed with artificial noises, and repeats such transmission over multiple rounds. Subsequently, in the next phase, each transmitter carefully sends a function of the net interference and artificial noises (generated in previous phase), which is simultaneously useful to all receivers. The equivocation analysis of the proposed scheme is non-trivial due to the repetition and retransmission strategies employed by the transmitters. This requires us to obtain a new result on the rank of product of two random non-square random matrices, which can be of independent interest. 


 \textbf{Organization of the paper:}  The rest of the paper is organized as follows. Section \ref{system_model_section}
describes the system models. The main results  and discussions are presented
in Section \ref{main_results}. Section \ref{seciton_proof_theorem} provides the achievable scheme under delayed CSIT and confidential messages. Sections \ref{appendix_e} and \ref{appendix_f}  discuss  two other secrecy constraints: 1) $K$-user interference channel with an external eavesdropper (IC-EE), and 2) $K$-user interference channel with confidential messages and an external eavesdropper (IC-CM-EE), respectively.
 We conclude the paper and discuss the future directions in Section \ref{conclusion}. Finally, the detailed proofs are deferred to the Appendices.

\textbf{Notations:} Boldface uppercase letters denote matrices (e.g., $\textbf{A}$),
boldface lowercase letters are used for vectors (e.g., $\textbf{a}$), we denote scalars by non-boldface lowercase letters (e.g., $x$), and sets by capital calligraphic letters (e.g., $\mathcal{X}$). The set of natural numbers, integer numbers, real numbers and complex numbers are denoted by $\mathds{N}$, $\mathds{Z}$, $\mathds{R}$ and $\mathds{C}$, respectively. For a general matrix $\mathbf{A}$ with dimensions of $M \times N$, $\textbf{A}^{T}$ and $\textbf{A}^{H}$ denote the transpose and Hermitian transpose of $\textbf{A}$, respectively. We denote the partitioned matrix of two matrices $\textbf{A}_{L \times N}$ and $\textbf{B}_{L \times M}$ as $(\textbf{A} : \textbf{B})$. We denote the identity matrix of the order $M$ with $\textbf{I}_{M}$. Let $h(\textbf{x})$ denote the differential entropy of a random vector $\textbf{x}$, and $I(\textbf{x}; \textbf{y})$ denote the mutual information between two random vectors $\textbf{x}$ and $\textbf{y}$. We denote a complex-Gaussian distribution with a mean $\mu$ and a variance $\sigma^{2}$
 by $\mathcal{CN}(\mu, \sigma^{2})$. For rounding operations on a variable $x$, we  use $\lfloor x \rfloor$ as the floor rounding operator on $x$ and $\lceil x \rceil$ as the ceiling rounding operator on $x$.

\begin{figure}
    \centering
        \includegraphics[scale=0.5]{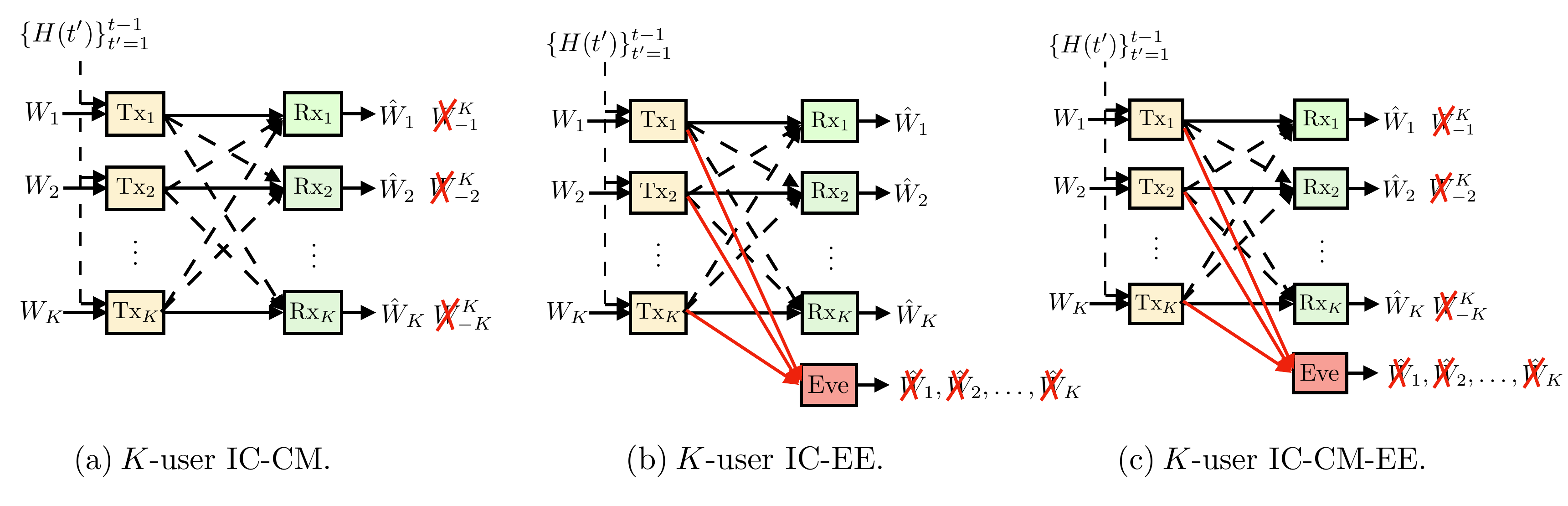}
    \caption{$K$-user  interference channel (IC)  with secrecy  and delayed CSIT. The model in (a) has confidential message (CM) constraints, model in (b) assumes the presence of an external eavesdropper (EE), and the model in (c) has both the secrecy constraints (CM $\&$ EE).}    
    \label{fig::fig_system_model}
\end{figure}
\section{System Model} \label{system_model_section}
 We consider the $K$-user interference channel with secrecy constraints and delayed CSIT (shown in Fig. \ref{fig::fig_system_model}). The input-output relationship at time slot $t$ is 
\begin{align}
y_{k}(t)  = h_{kk}(t) x_{k}(t) + \sum_{j=1, j\neq k}^{K}h_{kj}(t)x_{j}(t) + n_{k}(t),  \hspace{0.05in} z(t) & = \sum_{j=1}^{K} g_{j}(t)x_{j}(t) + n_{z}(t),
\end{align}
where $y_{k}(t)$ is the signal received at receiver $k$ at time $t$,  $h_{kj}(t) \sim \mathcal{CN}(0,1)$ is the channel coefficient at time $t$ between transmitter $j$ and receiver $k$, and  $x_{k}(t)$ is the transmitted signal from transmitter $k$ at time $t$ with an average power constraint $\mathds{E} \{|x_{k} (t)|^{2} \} \leq P$.   The additive noise $n_{k}(t)  \sim \mathcal{CN}(0,1) $  at receiver $k$ is also i.i.d. across users and time. $z(t)$ is the received signal at the eavesdropper at time $t$,  $g_{j}(t) \sim \mathcal{CN}(0,1)$ is the channel coefficient at time $t$ between transmitter $j$ and the external eavesdropper, and $n_{z}(t)  \sim \mathcal{CN}(0,1) $  is the additive noise at the  eavesdropper. The channel coefficients are assumed to be independent and identically distributed (i.i.d.) across time and users. We assume perfect CSI at all the receivers. We further assume that the CSIT is delayed, i.e., CSI  is available at each transmitter after one time slot without error. Also, we assume that the external eavesdropper's CSI is not available at the transmitters (i.e., no eavesdropper CSIT).

Let $R_{k}=\frac{\log_{2}\left(\left|\mathcal{W}_{k}\right|\right)}{\tau}$ denote the rate of message $W_{k}$ intended for receiver $k$, where $|\mathcal{W}_{k}|$ is the cardinality of the $k$th message.   A $(2^{\tau R_{1}}, 2^{\tau R_{2}},  \dots, 2^{\tau R_{K}}, \tau)$ code is  described by the set of encoding and decoding functions as follows:  the set of encoders at the transmitters are given as:
$\{\psi_{t}^{(k)}: {W}_{k} \times \{H(t')\}_{t'=1}^{t-1} \rightarrow x_{k}(t) \}_{t=1}^{\tau} , \forall k=1,\dots, K,$
where the message $W_{k}$  is uniformly distributed over the set $\mathcal{W}_{k}$, and $H(t') \triangleq \{h_{kj}(t')\}_{k =1, j =1}^{K}$ is the set of all channel gains at time $t'$. The transmitted signal from 
transmitter $k$ at time slot $t$ is given as:
$x_{k}(t)=\psi_{t} (W_{k}, \{H(t')\}_{t'=1}^{t-1})$. The decoding function at receiver $k$ is given by the following mapping:
$\phi^{(k)}: y_{k}^{(\tau)} \times  \{H(t)\}_{t=1}^{\tau} \rightarrow {W}_{k}$,
and the estimate of the message at receiver $k$ is defined as:
$\hat{W}_{k} = \phi^{(k)}(\{y_{k}(t), H(t)\}_{t=1}^{\tau})$. 
The rate tuple $(R_1,\ldots, R_K)$ is achievable if there exists a sequence of codes which satisfy the decodability constraints at the receivers, i.e.,
\begin{align}
\lim_{\tau \rightarrow \infty} \sup \text{Prob}\left[\hat{W}_{k} \neq  W_{k}\right] \leq \epsilon_{\tau}, \forall k = 1, \dots, K
\end{align}
and the corresponding secrecy requirement is satisfied. We consider three different secrecy requirements: 
\begin{enumerate}
\item IC-CM, Fig. (1a), all unintended messages are kept secure against each receiver, i.e., 
\begin{align}
 \lim_{\tau \rightarrow \infty} \sup \frac{1}{\tau} I\left(W_{-k}^{K}; y^{(\tau)}_{k} | W_{k}, \Omega \right) \leq  \epsilon_{\tau},  \forall k = 1, \dots, K, \label{conf_const_1}
\end{align}
\noindent where $\epsilon_{\tau} \rightarrow 0$ as $\tau \rightarrow \infty$,  $W_{-k}^{K} \triangleq  \{W_{1}, W_{2}, \dots, W_{K}\}  \backslash \{W_{k}\}$, and $\Omega \triangleq \{H(t)\}_{t=1}^{\tau}$ is the set of all channel gains over the channel uses.
\item IC-EE, Fig. (1b), all of the messages are kept secure against the external eavesdropper, i.e., 
\begin{align}
 \lim_{\tau \rightarrow \infty} \sup \frac{1}{\tau} I\left(W_{1}, W_{2}, \dots, W_{K}; z^{(\tau)}_{k} | \Omega \right) \leq  \epsilon_{\tau},  \forall k = 1, \dots, K. \label{eve_const_2}
\end{align}
\item IC-CM-EE, Fig. (1c), all of the messages are kept secure against both the $K-1$ unintended receivers and the external eavesdropper, i.e., we impose both secrecy constraints in (\ref{conf_const_1}) and (\ref{eve_const_2}).
\end{enumerate}

 \noindent The supremum of the achievable sum rate, $R_{s} \triangleq \sum_{k=1}^{K} R_{k}$,  is  defined as the secrecy sum capacity $C_{s}$.   The optimal sum secure degrees of freedom (SDoF) is then defined as follows:
\begin{equation}
\text{SDoF}^{*} \triangleq \lim_{P \rightarrow \infty} \frac{C_{S}}{\log\left(P\right)}.
\end{equation} 
{  SDoF represents the scaling of the secrecy capacity with $\log (P)$, where $P$ is the transmitted power, i.e., it is the pre-log factor of the secrecy capacity at high SNR.}

In the next Section, we present our main results on the achievable sum SDoF with the three different secrecy constraints  and delayed CSIT.

\section{Main Results} \label{main_results}

\begin{theorem}
\textit{For the $K$-user IC-CM with delayed CSIT, the following secure sum degrees of freedom is achievable:}
\label{theorem::1}
\end{theorem}
\begin{align}
 \text{SDoF}^{*}_\text{IC-CM} \geq \text{SDoF}_{\text{IC-CM}}^{\text{ach.}}=  \frac{K R (K-R-2)}{ (K-1) \times \left[R (R+1)+K\right]}  \label{sdof_equation_theorem}, 
\end{align}
\textit{where,} 
\begin{align}
R  = \bigg{\lfloor} \frac{-K+K \times \sqrt{1 +  \frac{(K-1)(K-2)}{K}}}{K-1} \bigg{\rfloor}.
\end{align}
{ We next simplify the above expression and present a lower bound on the achievable SDoF. Using this lower bound, we observe that the achievable SDoF scales with $\sqrt{K}$, where K is the number of users. 
\begin{corollary}\label{cor1}
\textit{ For the $K$-user IC-CM with delayed CSIT , the achievable SDoF in (\ref{sdof_equation_theorem}) is lower bounded as}
\end{corollary} 
\begin{align}
 \text{SDoF}_{\text{IC-CM}}^{\text{ach.}} > \frac{1}{2}  (\sqrt{K} -6)^{+}, \label{equation_coro}
\end{align}
\textit{where $(x)^{+}  \triangleq \max(x, 0)$}. \\
We present the proof of Theorem \ref{theorem::1} and Corollary \ref{cor1} in Section \ref{seciton_proof_theorem}.


\begin{theorem}
\textit{For the $K$-user IC-EE with delayed CSIT, the following secure sum degrees of freedom is achievable:}
\label{theorem::2}
\end{theorem}
\begin{align}
\text{SDoF}^{*}_\text{IC-EE} \geq  \text{SDoF}_{\text{IC-EE}}^{\text{ach.}} =  \frac{R (K-R-1)}{ R (R+1)+K}  \label{sdof_equation_theorem_2}, 
\end{align}
\textit{where,} 
\begin{align}
R  = \lfloor \sqrt{K} \rfloor - 1.
\end{align}
We next simplify the above expression and present a lower bound on the  $\text{SDoF}_{\text{sum}}^{\text{ach.}}$.
\begin{corollary}\label{cor3}
\textit{ For the $K$-user IC-EE with delayed CSIT, the achievable SDoF in (\ref{sdof_equation_theorem}) is lower bounded as}
\end{corollary} 
\begin{align}
 \text{SDoF}_{\text{IC-EE}}^{\text{ach.}} > \frac{1}{2}  (\sqrt{K} -3)^{+}. \label{equation_coro}
\end{align}
We present the proof of Theorem \ref{theorem::2} and Corollary \ref{cor3} in Section \ref{appendix_e}.
\begin{theorem}
\textit{For the $K$-user IC-CM-EE with delayed CSIT, the following secure sum degrees of freedom is achievable:}
\label{theorem::3}
\end{theorem}
\begin{align}
 \text{SDoF}^{*}_\text{IC-CM-EE}  \geq \text{SDoF}_{\text{IC-CM-EE}}^{\text{ach.}}=  \frac{K R (K-R-2)}{ (K-1) \times \left[R (R+1)+K\right]}  \label{sdof_equation_theorem_3}, 
\end{align}
\textit{where,} 
\begin{align}
R  = \bigg{\lfloor} \frac{-K+K \times \sqrt{1 +  \frac{(K-1)(K-2)}{K}}}{K-1} \bigg{\rfloor}.
\end{align}
In the next Corollary, we simplify the above expression and present a lower bound on the  $\text{SDoF}_{\text{sum}}^{\text{ach.}}$.
\begin{corollary}\label{cor4}
\textit{ For the $K$-user IC-CM-EE with delayed CSIT , the achievable SDoF in (\ref{sdof_equation_theorem}) is lower bounded as}
\end{corollary} 
\begin{align}
 \text{SDoF}_{\text{IC-CM-EE}}^{\text{ach.}} > \frac{1}{2}  (\sqrt{K} -6)^{+}. \label{equation_coro}
\end{align}
We present the proof of Theorem \ref{theorem::3} and Corollary \ref{cor4} in Section \ref{appendix_f}.

\noindent \textbf{Remark 1:} We next compare the secure sum DoF of the previous Theorems  to that of \cite{castanheira2017retrospective} (i.e., without secrecy constraints). For the $K$-user interference channel without secrecy constraints, the achievable sum DoF  in \cite{castanheira2017retrospective} is given as:
\begin{align}
 \text{DoF}^{\text{ach.}}_{\text{sum}}  &= \frac{K}{\lfloor \sqrt{K} \rfloor -1 + \frac{K}{\lfloor \sqrt{K} \rfloor}} 
\geq   \frac{\lfloor \sqrt{K} \rfloor}{2}  \overset{(a)} >  \frac{1}{2}(\sqrt{K}-1)^{+} \label{last_steps},
\end{align} 
where (a) follows from the fact that $\lfloor x \rfloor > x - 1$. Comparing these results together, we can conclude that the scaling behavior of the sum SDoF is still attainable and there is almost no scaling loss in sum SDoF compared to no secrecy case for sufficiently large $K$ (see Fig. \ref{fig::secure_comparison}).

\begin{figure}[t]
\centering
\includegraphics[scale=0.45]{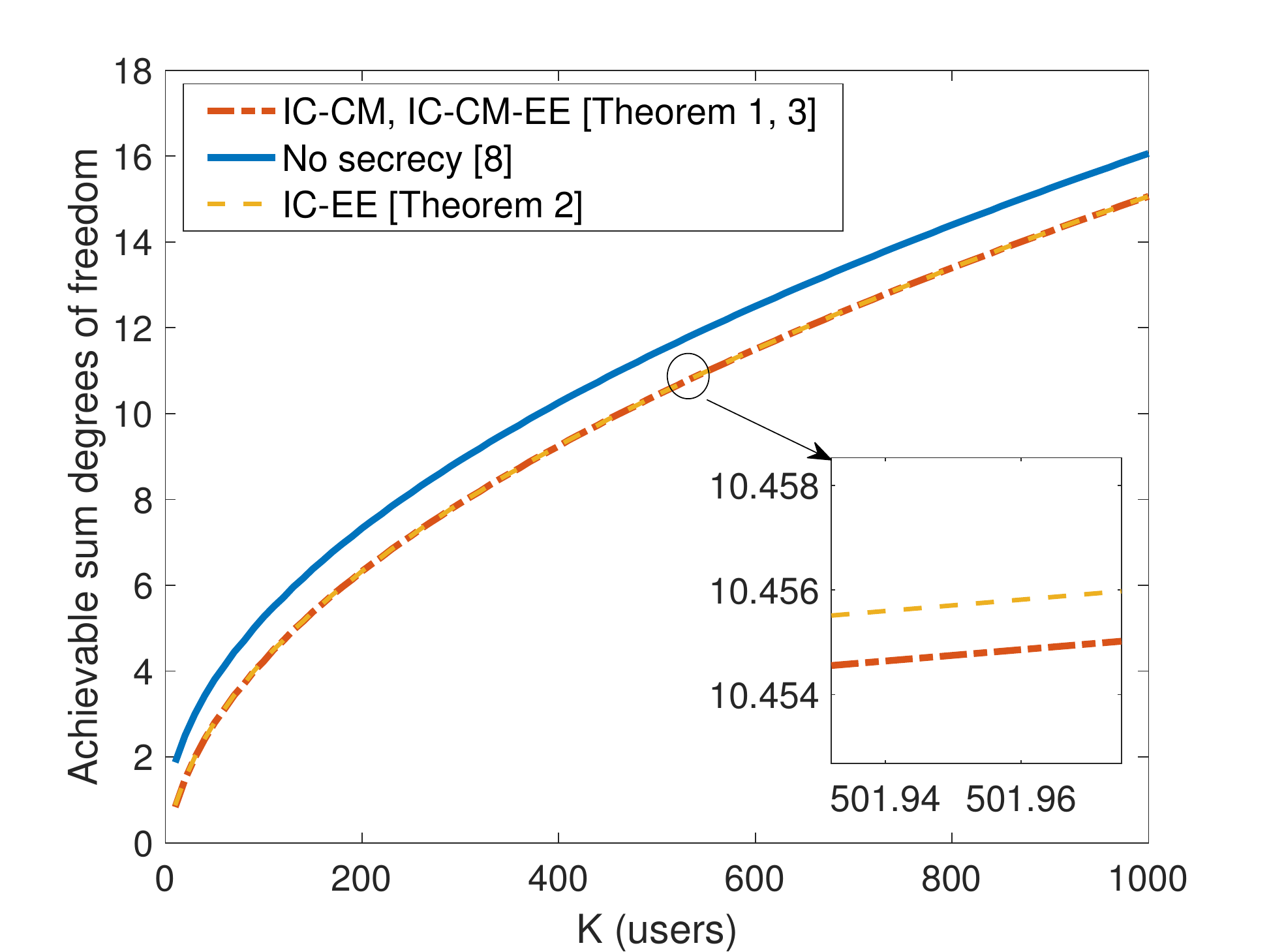}
\caption{Comparison of achievable DoF with delayed CSIT: with and without secrecy constraints.}
\label{fig::secure_comparison}
\end{figure}

\section{Proof of Theorem \ref{theorem::1}}\label{seciton_proof_theorem}

 \begin{figure}[t]
  \centering
\includegraphics[scale=0.5]{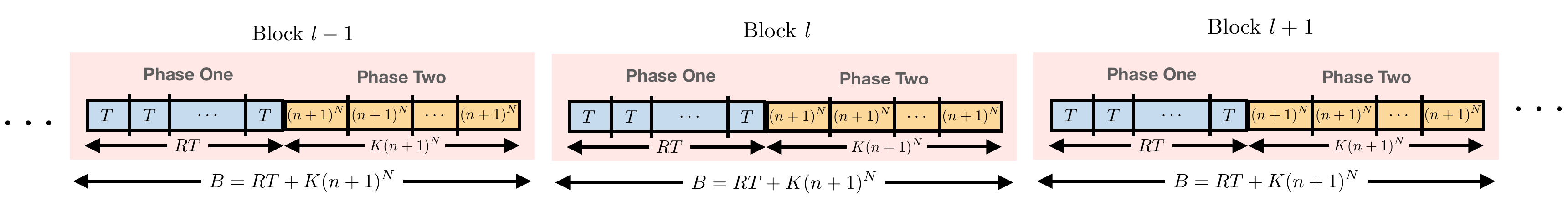}
\caption{Stochastic encoding over transmission blocks for our proposed scheme.}
\label{stochastic_encoding_block_editor}
\end{figure}

 In this Section, we present the proof of Theorem 1. The transmission scheme consists of $\tau$ transmission blocks, where each block is of duration $B$. Across blocks, we employ stochastic wiretap coding (similar to the techniques employed in the literature on compound wiretap channels, see \cite{xie2015secure, liang2009compound, liu2008discrete}). Within each block, the transmission is divided into two phases, which leverages delayed CSIT. In order to obtain the rate of the scheme, we first take the limit of number of blocks $\tau \rightarrow \infty$, followed by the limit $B \rightarrow \infty$. For a given block, if we denote the (B-length) input of transmitter $i$ as $\mathbf{x}_i$, and output of receiver $i$ as $\mathbf{y}_i$, then the secure rate achievable by stochastic wiretap coding is given by:
\begin{align}
R_{i} = \frac{ I(\mathbf{x}_{i}; \mathbf{y}_{i} | \Omega) - \max_{j \neq i} I(\mathbf{x}_{i}; \mathbf{y}_{j}  | \Omega  )} {B}, \hspace{0.05in} i = 1,2, \dots, K.
\end{align} 
Fig. \ref{stochastic_encoding_block_editor} gives an overview for these steps: stochastic encoding over blocks, and the two-phase scheme within each block that leverages delayed CSIT.

\subsection*{Overview of the achievability scheme and SDoF analysis}
In this subsection, we present our secure transmission scheme. We consider a transmission block of length $RT  + K (n+1)^{N}=R ( R n^{N} + (n+1)^{N}) + K (n+1)^{N}$, where $R$ denotes the number of transmission rounds and $N = R K (K-1)$, and $n$ is an integer.  The transmission scheme works over two phases. 
The goal of each transmitter is to securely send $T_{1} = Rn^{N} + (n+1)^{N} - \lceil \frac{RT + (K-1) (n+1)^{N}}{K-1}\rceil$ information symbols to its corresponding receiver. In the first phase, each transmitter sends random linear combinations of the $T_{1}$ information symbols and the  $T_{2} = \lceil \frac{RT + (K-1) (n+1)^{N}}{K-1}\rceil$ artificial noise symbols  in $T$ time slots. Each transmitter repeats such transmission for $R$ rounds, and hence, phase one spans $RT$ time slots.

By the end of phase one, each receiver applies local interference alignment on its received signal to reduce the dimension of the aggregate interference. In the second phase, each transmitter knows the channel coefficients of phase one due to delayed CSIT. Subsequently, each transmitter sends a function of the net interference and artificial noises  (generated in previous phase) which is simultaneously useful to all receivers. More specifically, each transmitter seperately sends $(n+1)^{N}$ linear equations of the past interference to all receivers. Therefore, phase $2$ spans $K (n+1)^{N}$ time slots. 

By the end of both phases, each receiver is able to decode its desired $T_{1}$ information symbols while satisfying the confidentiality constraints. The main aspect is that the  parameters of the scheme (i.e., number of artificial noise symbols, number of repetition rounds and durations of the phases)  must be carefully selected to allow for reliable decoding of legitimate symbols, while satisfying the confidentiality constraints. 

Therefore, the transmission scheme spans $RT  + K (n+1)^{N}$ time slots,   this scheme leads to the following achievable SDoF:
\begin{align}
 \text{SDoF}_{\text{IC-CM}}^{\text{ach.}}  = \frac{K R (K-R-2)}{ (K-1) \times \left[R (R+1)+K\right]}.  \label{approximated}
\end{align}
We  calculate the achievable sum SDoF of this scheme in full detail in subsection \ref{rate_analysis_CM}. Before we present the details of the scheme, we first optimize the achievable SDoF in (\ref{approximated}) with respect to the number of rounds $R$ and also simplify the above expression, which leads to the expression in Corollary \ref{cor1}.
\begin{lemma} \label{concavity_proof}
\textit{The optimal value of $R^{*}$ which maximizes  (\ref{approximated}) is given by}
\begin{align}
R^{*}=  \bigg{\lfloor} \frac{-K+K \times \sqrt{1 +  \frac{(K-1)(K-2)}{K}}}{K-1} \bigg{\rfloor}.
\end{align}
\end{lemma}
Now, in order simplify the obtained  expression in  (\ref{approximated}), we state the following Corollary.
\begin{corollary}\label{cor1_proof}
\textit{The optimal value of number of rounds $R^{*}$ is lower bounded by}
\begin{align}
R^{*} > \sqrt{K} - 5 = R_{\text{lb}}.
\end{align}
\end{corollary} 
We present the proof of Lemma \ref{concavity_proof} and Corollary \ref{cor1_proof} in Appendix \ref{section_a}.
\begin{figure}
\centering
\includegraphics[scale=0.45]{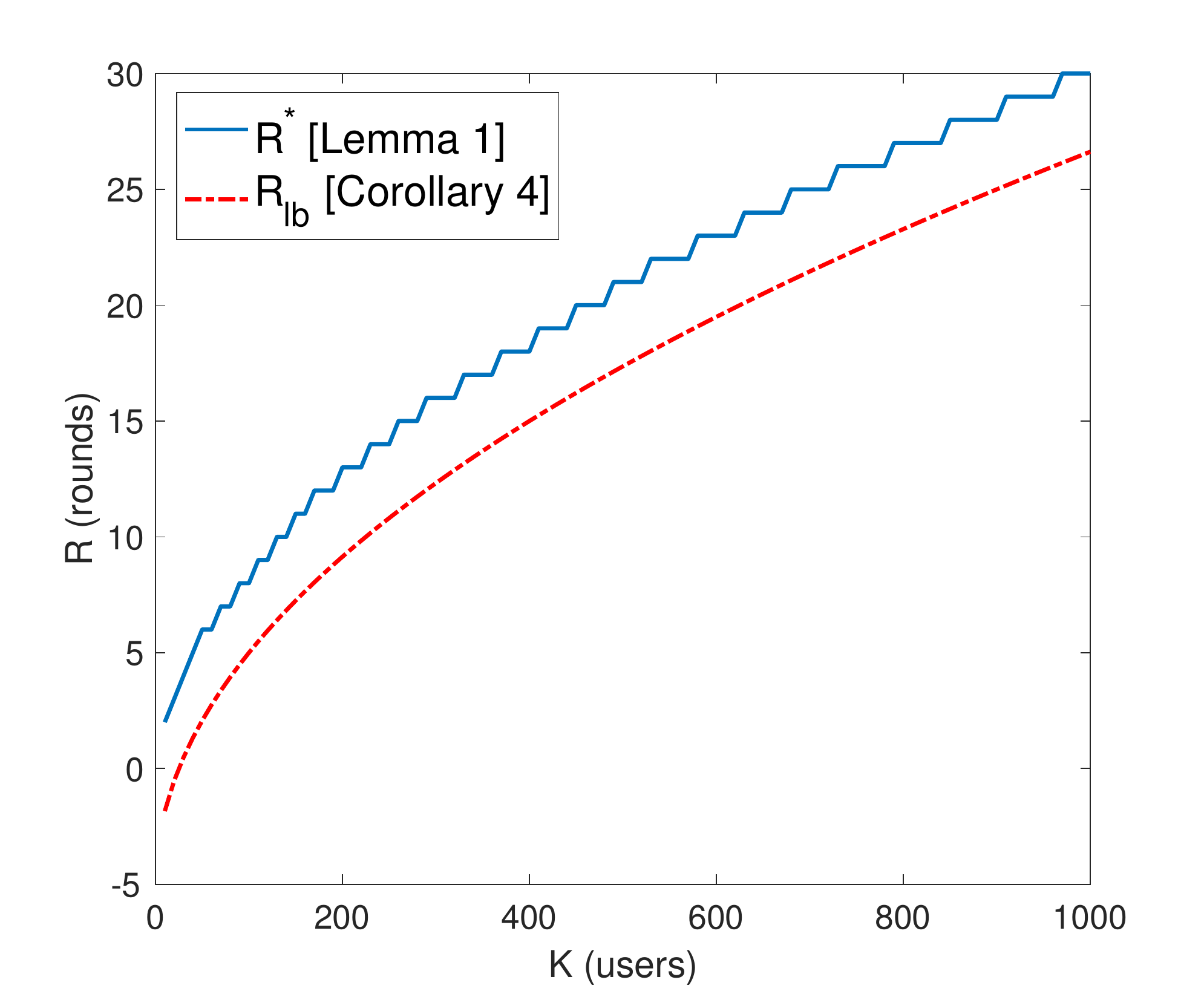}
\caption{Comparison between the optimal value of $R$ (number of rounds in phase one of the scheme) and its lower bound.}
\label{fig::approximation}
\end{figure}
\noindent Fig. \ref{fig::approximation} depicts a comparison of between the two values of $R$ (i.e., optimal $R^{*}$ and lower bound $R_{\text{lb}}$). 
By substituting  $R_{\text{lb}}$ in (\ref{approximated}) leads to a lower bound on  $\text{SDoF}_{\text{IC-CM}}^{\text{ach.}}$ as follows:
\begin{align}
 \text{SDoF}_{\text{IC-CM}}^{\text{ach.}} & =   \frac{K R (K-R-2)}{ (K-1) \times \left[R (R+1)+K\right]},\\
&>  \frac{R (K-R-2)}{ R (R+1)+K} = \frac{(\sqrt{K} -5) (K - \sqrt{K} + 3)}{(\sqrt{K} -5) (\sqrt{K} -4) + K }, \\
&  \overset{(a)}  =  \frac{K \sqrt{K} - 6K + 8 \sqrt{K} - 15}{2 K - 9 \sqrt{K} +20}, \label{term_before} \\
&  > \frac{K \sqrt{K} - 6K + 8 \sqrt{K} - 15}{2 K},  \label{term_2} \\
&  =  \frac{\sqrt{K} - 6}{2 } + \frac{8 \sqrt{K} - 15}{ 2K}, \\ 
& \overset {(b)} >  \frac{1}{2} (\sqrt{K}-6)^{+},   \label{term_5}
 \end{align}
where in (a), the term $ -9 \sqrt{K} + 20$ in the  denominator is negative, $\forall K \geq 5$ , so neglecting this term gives us (\ref{term_2}). In step (b), since the term $8 \sqrt{K} - 15$ is  positive, $\forall K \geq 4$, hence omitting this term gives  (\ref{term_5}).  To this end, we get (\ref{term_5}) which shows the scaling of the achievable SDoF with $K$, the number of users.

\subsection{Detailed description of the achievability scheme}

\begin{figure*}[h]
  \centering
\includegraphics[scale=0.35]{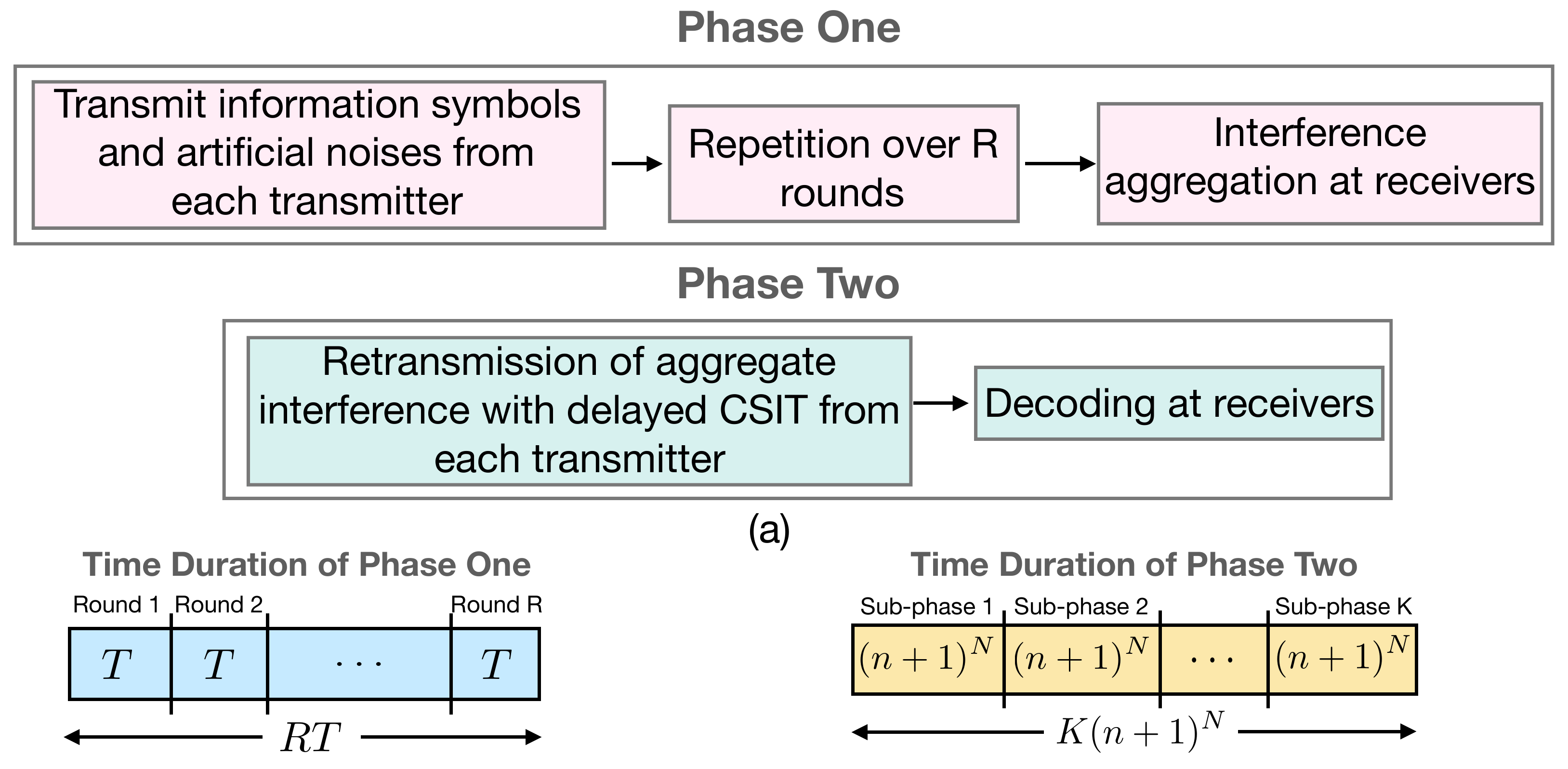}
\caption{(a) Block diagram for the transmission scheme and (b) Time duration of the phases.}
\label{fig::scheme_flow}
\end{figure*}

\noindent Fig. \ref{fig::scheme_flow} depicts an overview of the two transmission phases. 
We now present the transmission scheme in full detail. For our scheme, we collectively denote the $L$ symbols transmitted over $L$ time slots as a super symbol and call this as the $L$ symbol extension of the channel.  
With the extended channel, the signal vector at the $k$th receiver can be expressed as  
\begin{align}
\mathbf{y}_{k} = \sum_{j=1}^{K}\mathbf{H}_{kj} \mathbf{x}_{j}+ \mathbf{n}_{k},
\end{align}
where $\mathbf{x}_{j}$ is a $L \times 1$ column vector representing the $L$ symbols transmitted by transmitter $k$ in $L$ time slots.  $\mathbf{H}_{kj}$ is a  $L \times L$ diagonal matrix representing the $L$ symbol extension of the channel as follows:
\begin{align}
\mathbf{H}_{kj} = \text{diag}\left(h_{kj}(1), h_{kj}(2), \dots, h_{kj}(L) \right),
\end{align}
where $h_{kj}(t)$ is the channel coefficient between transmitter $j$ and receiver $k$ at time slot $t$. Now we proceed to the proposed scheme which works over two phases.
\begin{figure*}[h]
\centering
  \includegraphics[scale=0.45]{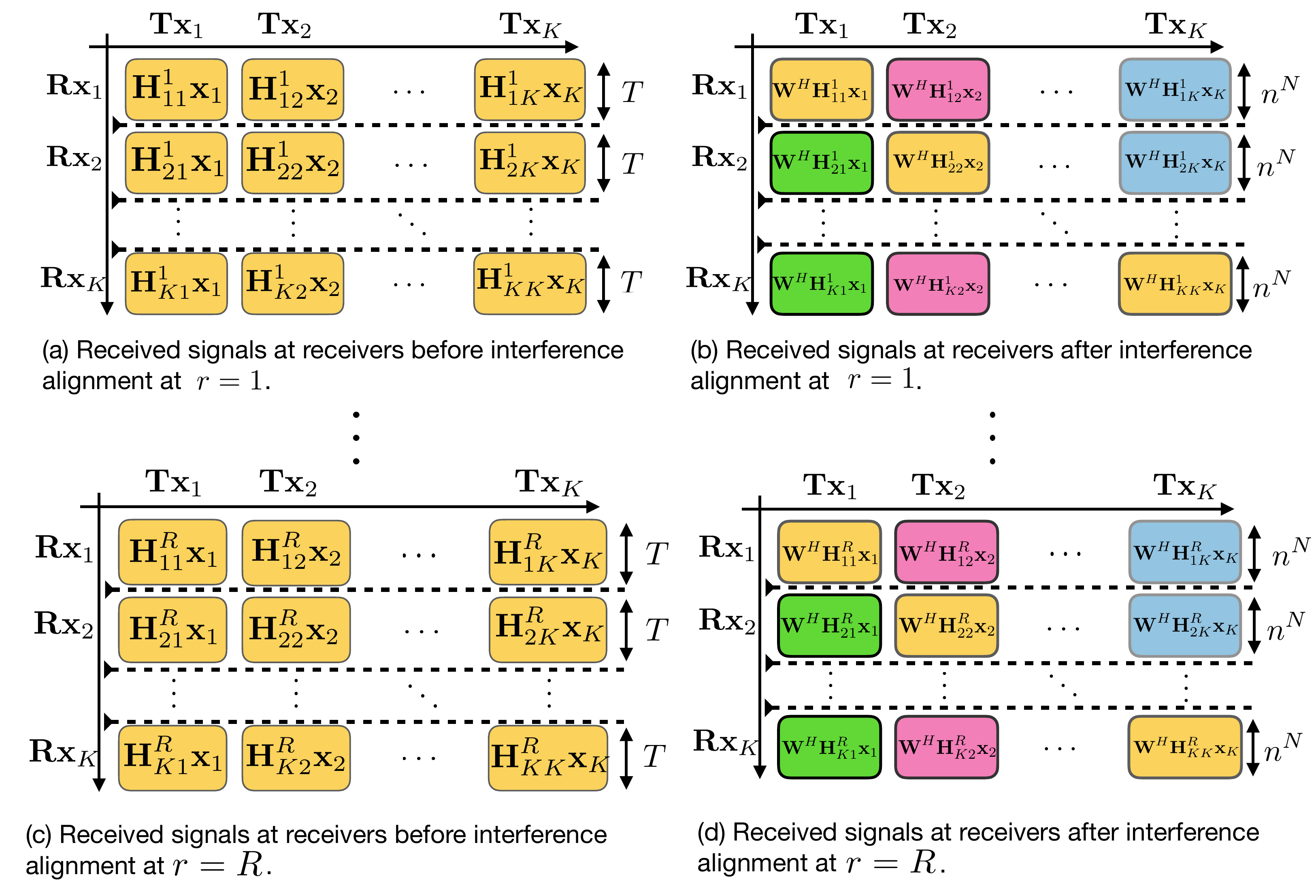}
\caption{Graphical representation for the first phase of the proposed scheme.}
\label{fig:test}
\end{figure*}

\subsection*{Phase 1: Interference creation with information symbols and artificial noises:}

 Recall that the goal of each transmitter is to send $T_{1}$ information symbols securely to its respective receiver. This phase is comprised of $R$ rounds, where, in each round, every  transmitter $j$ sends  linear combinations of the $T_{1}$ information symbols  $\mathbf{s}_{j} \in \mathds{C}^{T_{1} \times 1}$,  mixed with $T_{2}$ artificial noises $\mathbf{u}_{j} \in \mathds{C}^{T_{2} \times 1}$,  where the elements of $\mathbf{u}_{j}$ are drawn from complex-Gaussian distribution with average power $P$.  Hence, the  signal sent by  transmitter $j$ in each round $r$ can be written as
\begin{align}
\mathbf{x}_{j} = \mathbf{V}_{j} \left[ \begin{matrix}  \mathbf{s}_{j} \\  \mathbf{u}_{j} \end{matrix} \right], \hspace{0.05in} \forall j = 1, 2, \dots, K, 
\end{align}
where  $\mathbf{V}_{j},  \forall j =1, 2, \dots, K$ is a random \textit{mixing} matrix of dimension $T \times T$ whose elements are drawn from complex-Gaussian distribution with zero mean and unit variance at transmitter $j$. $\mathbf{V}_{j}, \forall j =1, 2, \dots, K$ is known at all terminals (all transmitters and receivers). The received signal at receiver $k$ for round $r \in \{1, 2, \dots, R\}$ is given by
\begin{align}
\mathbf{y}_{k}^{r} = \sum_{j=1}^{K}\mathbf{H}_{kj}^{r} \mathbf{x}_{j} + \mathbf{n}_{k}^{r},
\end{align}
where $\mathbf{x}_{j}$ is the $T \times 1$ column vector representing the $T$ symbol extension of the transmitted symbols from transmitter $j$, and $\mathbf{n}_{k}^{r}$ represents the  receiver noise in round $r$ at receiver $k$.  This phase spans $RT$ time slots where $R \in \mathds{N}$  is the number of transmission rounds and $T = Rn^{N} + (n+1)^{N}$ time slots where $n \in \mathds{N}$ and $N = R K (K-1)$.  
\subsection*{Interference aggregation at receivers}
At the end of phase $1$, each receiver $k$ has the signals $ \mathbf{y}_{k} = \{\mathbf{y}_{k}^{r}\}_{r=1}^{R}$, over $R$ rounds. Each receiver performs a linear post-processing of its received signals in order to align the aggregate interference (generated from symbols and artificial noises) from the $(K-1)$ unintended transmitters. In particular, each receiver multiplies its received signals in the $r$th block with a matrix $\mathbf{W}$ (of dimension $T \times n^{N}$) as follows:
\begin{align}
\tilde{\mathbf{y}}_{k}^{r} &= \mathbf{W}^{H} \mathbf{y}_{k}^{r} = \mathbf{W}^{H} \bigg{ (}\sum_{j=1}^{K} \mathbf{H}_{kj}^{r} \mathbf{x}_{j} + \mathbf{n}_{k}^{r} \bigg{)},\\
&=\mathbf{W}^{H} \mathbf{H}_{kk}^{r} \mathbf{x}_{k} +  \sum_{j \neq k} \mathbf{W}^{H}  \mathbf{H}_{kj}^{r} \mathbf{x}_{j} + \mathbf{W}^{H} \mathbf{n}_{k}^{r},\\ 
&=\mathbf{W}^{H}  \mathbf{H}_{kk}^{r} \mathbf{x}_{k} +   \sum_{j \neq k} \mathbf{W}^{H} \mathbf{H}_{kj}^{r} \mathbf{x}_{j} +  \tilde{\mathbf{n}}_{k}^{r}.
\end{align}
The goal is to design the matrices $\mathbf{W}$ and  $\mathbf{X}$ such that
\begin{align}
\mathbf{W}^{H} \mathbf{H}_{kj}^{r} \prec \mathbf{X}, \forall k =1,2, \dots, K,  k \neq j,  \forall r = 1,2, \dots, R, \label{alpha_equation}
\end{align}
where $\mathbf{X} \in \mathds{C}^{(n+1)^{N} \times T}$.  Here the notation $\mathbf{A} \prec \mathbf{B}$ means that the set of row vectors of matrix $\mathbf{A}$ is a subset of the row vectors of matrix $\mathbf{B}$. To this end,  we choose $\mathbf{W}$ and $\mathbf{X}$  as follows:
\begin{align}
\mathbf{W} = \left[ \prod_{(r,m,i) \in \mathcal{S}} (\mathbf{H}_{mi}^{r^{(n_{mi}^{r})}})^{H} \mathds{1}: 0 \leq n_{mi}^{r} \leq n-1 \right],\\
\mathbf{X} = \left[ \prod_{(r,m,i) \in \mathcal{S}} (\mathbf{H}_{mi}^{r^{(n_{mi}^{r})}})^{H} \mathds{1}: 0 \leq n_{mi}^{r} \leq n \right]^{H},
\end{align}
where $\mathds{1}$ is the all ones column vector and the set $\mathcal{S} = \{(r, m, i): \forall  r \in\{1, \dots, R\}, \forall m \neq i \in \{1, \dots, K\}\}$. Note that the set  $\mathcal{S} $  does not contain the channel  matrix $\mathbf{H}_{kk}^{r}$ that carries the information symbols intended to receiver $k$. However,  multiplying with any channel gain that appears in $\mathbf{W}$ results in aligning this signal within $\mathbf{X}$ asymptotically.   It is worth noting that, $\mathbf{X}$ defines all the possible interference generated by the transmitters at all receivers. Hence, this choice of $\mathbf{X}$ and $\mathbf{W}$ guarantees that the alignment condition (\ref{alpha_equation}) is satisfied.
Therefore, the received signal after  post-processing  in  round $r$ at receiver $k$  can be written as
\begin{align}
\tilde{\mathbf{y}}_{k}^{r}& = \mathbf{W}^{H} \mathbf{H}_{kk}^{r} \mathbf{x}_{k} + \sum_{ j\neq k} \mathbf{W}^{H} \mathbf{H}_{kj}^{r} \mathbf{x}_{j}   + \mathbf{W}^{H} \mathbf{n}_{k}^{r},\\\
&=  \mathbf{W}^{H} \mathbf{H}_{kk}^{r} \mathbf{x}_{k}  + \sum_{ j\neq k}  \Pi_{kj}^{r} \mathbf{X} \mathbf{x}_{j} +  \mathbf{W}^{H} \mathbf{n}_{k}^{r},
\end{align}
where $\Pi_{kj}^{r} \in \mathds{C}^{n^{N} \times (n+1)^{N}}$ is a selection and permutation matrix. Now after the end of phase 1, receiver $k$ has $Rn^{N}$ equations of $T$ desired symbols (which are composed of $T_{1}$ information symbols and $T_{2}$ artificial noises generated by the transmitter $k$) plus $(K-1)$ interference terms, which are of dimension $(n+1)^{N}$. Fig. \ref{fig:test} gives a detailed structure for the first phase of the transmission scheme.

\begin{figure*}[h]
  \centering
  \includegraphics[scale=0.45]{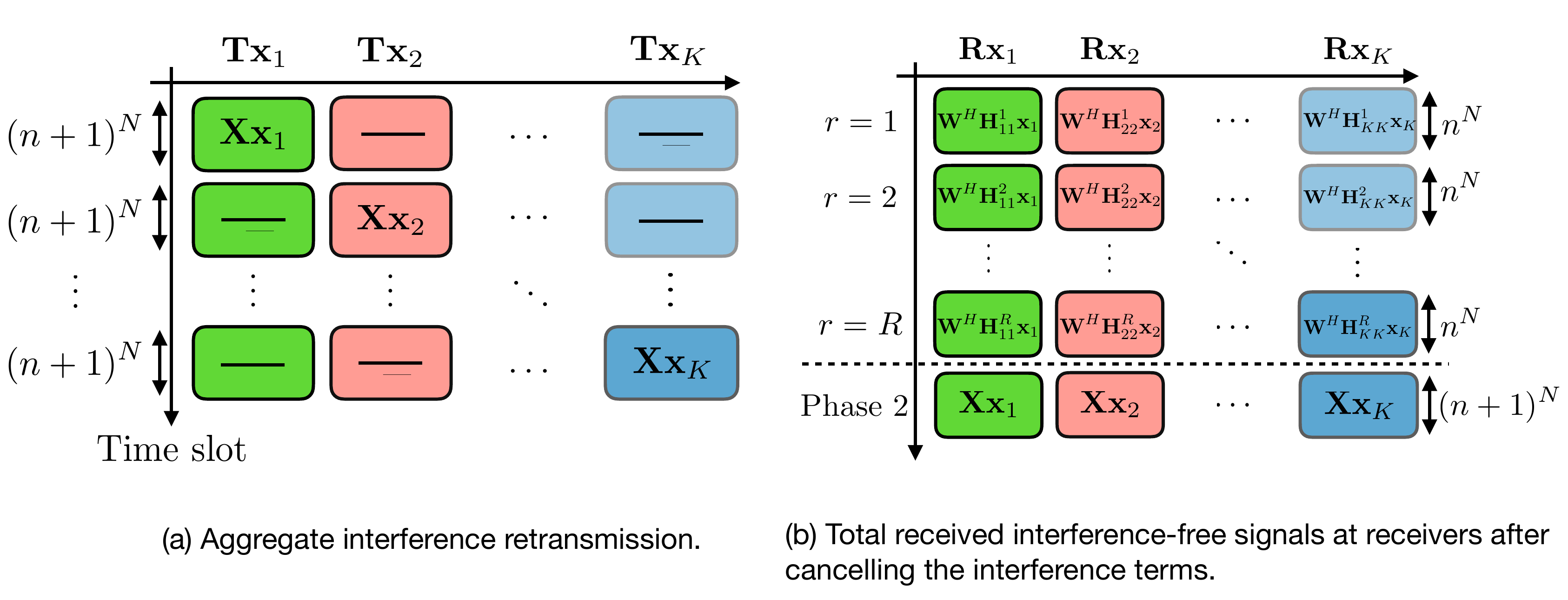}
\caption{Graphical representation for the second phase of the proposed scheme.}
\label{phase_two_block}
\end{figure*}

\subsection*{Phase 2: Re-transmission of aggregate interference with delayed CSIT:}
For the second phase, each transmitter $k$ uses $(n+1)^{N}$ time slots to re-transmit the aggregated interference  $(\mathbf{X} \mathbf{x}_k)$ generated in the first phase at the receivers, which is sufficient to cancel out the interference term at receiver $j \neq k$, and to provide additional $(n+1)^{N}$ equations of the desired symbols to receiver $k$. Then, this phase spans $K (n+1)^{N}$ time slots. The transmitted signal from transmitter $k$ is as follows:
\begin{align}
\mathbf{z}_{k} =  \mathbf{X} \hspace{0.05in} \mathbf{x}_{k}, \forall k= 1, 2, \dots, K.
\end{align}
\subsection*{Decoding at receivers:}
At the end of phase $2$, the interference at receiver $k$ is removed by subtracting the terms \\
 $\sum_{j=1, j\neq k}  \Pi_{kj}^{r} \mathbf{X} \hspace{0.05in}  \mathbf{x}_{j}$ from the equalized signal $\tilde{\mathbf{y}}_{k}^{r}$, i.e., (ignoring the additive noise $\mathbf{n}_{k}^{r}$)
\begin{align}
 \mathbf{W}^{H} \mathbf{H}_{kk}^{r} \mathbf{x}_{k}  = \tilde{\mathbf{y}}_{k}^{r} -  \sum_{j=1, j\neq k}  \Pi_{kj}^{r} \mathbf{X} \hspace{0.05in}\mathbf{x}_{j}.
\end{align}
Canceling the interference terms leaves each receiver $k, \forall k \in \{1, \dots, K\}$ with $R n ^{N}$ useful linear equations besides $(n+1)^{N}$ useful equations from transmitter $k$ (from phase $2$). At the end of phase $2$, receiver $k$ will collectively get the following signal,
\begin{align}
 {\underbrace {\left[\mathbf{X}^{H}, (\mathbf{W}^{H}\mathbf{H}_{kk}^{1})^{H}, \dots, (\mathbf{W}^{H}\mathbf{H}_{kk}^{R})^{H}\right]}_{\mathbf{B}_{k}} }^{H} \mathbf{V}_{k} \left[ \begin{matrix}  \mathbf{s}_{k} \\  \mathbf{u}_{k} \end{matrix} \right].  \label{important_equation}
\end{align} 
Therefore, at the end of phase $2$, each receiver has enough linear equations of the desired symbols. In order to ensure decodability, we need to prove that   $\mathbf{B}_{k} \mathbf{V}_{k}$ is full rank and hence each receiver will be able to decode its desired $T_{1}$ information symbols. First, we notice that $\mathbf{V}_{k}$ is full rank matrix and hence $\text{rank}(\mathbf{B}_{k} \mathbf{V}_{k}) = \text{rank}(\mathbf{B}_{k})$ \cite{horn2012matrix}. In Appendix \ref{linear_indpendence_proof}, we show that  $\mathbf{B}_{k}$ is full rank. Fig. \ref{phase_two_block} gives a detailed structure for the second phase of the transmission scheme. 

Before we start the achievable secure rate analysis, we want to highlight first on the dimensions of the information symbols  $\mathbf{s}_{i} \in \mathds{C}^{T_{1} \times 1}$ and the artificial noises  $\mathbf{u}_{i} \in \mathds{C}^{T_{2} \times 1}$, $\forall i = 1, 2, \dots, K$. 

\subsection*{Choice of $T_{1}$ and $T_{2}$ to satisfy the confidentiality constraints:}

Without loss of generality, let us consider receiver $1$.  After decoding $\mathbf{s}_{1}$ and $\mathbf{u}_{1}$,  receiver $1$ will have $RT$ equations of $\{\mathbf{s}_{i}\}_{i=2}^{K}$,  $\{\mathbf{u}_{i}\}_{i=2}^{K}$ from phase one, and $(K-1)(n+1)^{N}$ equations of $\{\mathbf{s}_{i}\}_{i=2}^{K}$,  $\{\mathbf{u}_{i}\}_{i=2}^{K}$ from phase two.  Then, the total number of equations seen at receiver $1$ is $RT + (K-1) (n+1)^{N}$. Hence, in order to keep the unintended information symbols of $(K-1)$ transmitters at this receiver secure, we require that the number of these equations must be at most equal to the total number of the artificial noise dimensions of the $(K-1)$ transmitters, i.e.,
\begin{align}
RT + (K-1) (n+1)^{N} \leq (K-1) T_{2}.
\end{align}
Therefore, we choose $T_{2}$ as
\begin{align}
T_{2} =  \bigg{\lceil} \frac{RT + (K-1) (n+1)^{N} }{K-1} \bigg{\rceil}.  \label{mind}
\end{align}
 Note that since $T = T_{1} + T_{2}$, so we can get $T_{1}$  as follows:
\begin{align}
 T_{1} = Rn^{N} + (n+1)^{N} -  \bigg{\lceil} \frac{RT + (K-1) (n+1)^{N}}{K-1} \bigg{\rceil}.
\end{align} 
We  next  compute the achievable secrecy rates and SDoF for the $K$-user interference channel with confidential messages and delayed CSIT.

\subsection{Secrecy Rate and SDoF Calculation} \label{rate_analysis_CM}

 Using stochastic encoding described in Appendix XII, for a block length $B = RT + K(n+1)^{N}$, the following secure rate is achievable: 

\begin{align}
R_{i} = \frac{ I(\mathbf{x}_{i}; \mathbf{y}_{i} | \Omega) - \max_{j \neq i} I(\mathbf{x}_{i}; \mathbf{y}_{j}  | \Omega  )} {R T + K (n+1)^{N}}, \hspace{0.05in} i = 1,2, \dots, K,  \label{korner_rate}
\end{align} 
where $I(\mathbf{x}_{i}; \mathbf{y}_{i} | \Omega)$ is the mutual information between the transmitted  symbols vector $\mathbf{x}_{i}$ and $\mathbf{y}_{i}$,  the received composite signal vector at the intended receiver $i$, given the knowledge of the channel coefficients.  $I(\mathbf{x}_{i}; \mathbf{y}_{j} | \Omega)$ is the mutual information between $\mathbf{x}_{i}$ and $\mathbf{y}_{j}$, the received composite signal vector at the unintended receiver $j$, i.e., the strongest adversary with respect to transmitter $i$. In terms of differential entropy, we can write, 
\begin{align}
 \textbf{Term A:} \hspace{0.1in} I(\mathbf{x}_{i}; \mathbf{y}_{i} | \Omega)  = h(\mathbf{y}_{i} | \Omega ) - h(\mathbf{y}_{i} | \mathbf{x}_{i}, \Omega), \hspace{0.05in} i=1,2, \dots, K,  \\
  \textbf{Term B:} \hspace{0.1in} I(\mathbf{x}_{i}; \mathbf{y}_{j} | \Omega)  = h(\mathbf{y}_{j} | \Omega ) - h(\mathbf{y}_{j} | \mathbf{x}_{i}, \Omega), \hspace{0.05in} j=1, 2, \dots, K, j \neq i. 
\end{align}

We collectively write the received signal $\mathbf{y}_{i}$ at receiver $i$ over $RT+ K (n+1)^{N}$ time slots as follows:
\begin{align}
\mathbf{y}_{i} = \mathbf{A}_{i} \mathbf{V} \mathbf{q} + \mathbf{n}_{i}, \forall i =1,2, \dots, K, 
\end{align}
where,

\begin{align}
 &\mathbf{A}_{i} =  \begin{bmatrix}
\mathbf{C}_{i} \\ \hline  \mathbf{D}_{i}
\end{bmatrix}, \hspace{0.05in} \mathbf{C}_{i} =  \left[ \begin{matrix} \mathbf{H}_{i1}^{1} & \mathbf{H}_{i2}^{1} & \cdots  & \mathbf{H}_{iK}^{1} \\ \mathbf{H}_{i1}^{2} & \mathbf{H}_{i2}^{2} & \cdots  & \mathbf{H}_{iK}^{2} \\ \vdots  & \vdots  & \cdots  & \vdots \\ \mathbf{H}_{i1}^{R} & \mathbf{H}_{i2}^{R} & \cdots  & \mathbf{H}_{iK}^{R} \end{matrix} \right], \nonumber \\ 
&\mathbf{D}_{i} = \text{blkdiag}(\tilde{\mathbf{H}}_{i1} \mathbf{X}, \dots, \tilde{\mathbf{H}}_{iK} \mathbf{X}).
\end{align}

\noindent where  $\mathbf{A}_{i}$ has dimensions of $(RT + K (n+1)^{N}) \times KT$. Note that  $\mathbf{A}_{i}$ is partitioned into two sub matrices $\mathbf{C}_{i}$ and $\mathbf{D}_{i}$.  $\mathbf{C}_{i}$ consists of block matrices, where each block matrix has dimensions of $T \times T$ whose elements are i.i.d. drawn from a continuous distribution and hence, it is full rank, almost surely (i.e., $\text{rank}(\mathbf{C}_{i}) = RT$).  $\mathbf{D}_{i}$ has a block diagonal structure (each block matrix has dimensions of $(n+1)^{N} \times T$) since the transmission in phase two of the scheme is done in TDMA fashion. Note that each block is a full rank matrix (i.e., $\text{rank} (\tilde{\mathbf{H}}_{ij} \mathbf{X}) = \text{rank}(\mathbf{X}) = (n+1)^{N}, \forall j =1, \dots, K$).  The matrix $\mathbf{X}$ is a full rank matrix as proved in \cite{seif2018arxiv}. The matrix $\mathbf{V}$ can be written as follows:
\begin{align}
\mathbf{V} = \text{blkdiag} (\mathbf{V}_{1}, \mathbf{V}_{2}, \dots, \mathbf{V}_{K}),
\end{align}
where $\mathbf{V}$ is the block diagonal matrix with dimensions of $KT \times K T$. Furthermore, we  write
\begin{align}
\mathbf{q} = \left[ \begin{matrix} \mathbf{s}_{1}^{T} & \mathbf{u}_{1}^{T} & \mathbf{s}_{2}^{T} & \mathbf{u}_{2}^{T} & \cdots  & \mathbf{s}_{K}^{T} & \mathbf{u}_{K}^{T} \end{matrix} \right]^{T}
\end{align}
as a column vector of length $ K T$, which contains the information symbols and the artificial noises of transmitters $1, \dots, K$.

Before we proceed, we present two Lemmas which are proved in Appendices X and XI.
\begin{lemma} \label{lemma_one}
 \textit{Let $\mathbf{A}$ be a matrix with dimension $M \times N$ and $\mathbf{X}=(x_{1}, \dots ,x_{N})^{T}$ be a zero-mean  jointly complex Gaussian random vector with  covariance matrix $P\mathbf{I}$. Also, let $\mathbf{N}=(n_{1}, \dots, n_{M})^{T}$ be a  zero-mean jointly complex Gaussian random vector with covariance matrix $\sigma^{2}\mathbf{I}$, independent of $\mathbf{X}$, then} 
\end{lemma}
\begin{align}
h(\mathbf{A} \mathbf{X} + \mathbf{N}) = \log(\pi e)^{M} + \sum_{i = 1}^{\text{rank} (\mathbf{A})} \log(\lambda_{i} P + \sigma^{2}).
\end{align}
\textit{where $\{\lambda_{i}\}_{i=1}^{\text{rank}(\mathbf{A})}$ are the singular values of $\mathbf{A}$.}


\begin{lemma} \label{lemma_2}
\textit{Consider two matrices $\mathbf{A}_{M \times N}$ and $\mathbf{B}_{N \times M}$ where $M \leq N$. The elements of matrix $\mathbf{B}$ are chosen independently from the entries of $\mathbf{A}$ at random from a continuous distribution. Then, }
\end{lemma}
\begin{equation}
\text{rank} (\mathbf{A}\mathbf{B}) = \text{rank} (\mathbf{A}), \hspace{0.05 in} \textit{almost surely}.
\end{equation}

Without loss of generality, let us consider the first transmitter. The received signal at the first receiver after removing the $(K-1)$ interference terms is written as follows: 
\begin{align}
\tilde{\mathbf{y}}_{1}= \underbrace{ \left[ \begin{matrix}
\mathbf{W}^{H}  & \mathbf{0} & \dots  & \mathbf{0} & \mathbf{0} \\ \mathbf{0} & \mathbf{W}^{H} & \dots  & \mathbf{0} & \mathbf{0} \\ \vdots & \vdots & \ddots & \vdots & \vdots \\ \mathbf{0} & \mathbf{0} & \dots & \mathbf{W}^{H} & \mathbf{0} \\ \mathbf{0} & \mathbf{0} & \dots & \mathbf{0} & \mathbf{I}
\end{matrix} \right]}_{\mathbf{\Psi}}  \left( \underbrace{\left[ \begin{matrix} \mathbf{H}_{11}^{1} \\ \mathbf{H}_{11}^{2}  \\ \vdots \\ \mathbf{H}_{11}^{R} \\ \tilde{\mathbf{H}}_{11} \mathbf{X} \end{matrix} \right] \mathbf{V}_{1}}_{\mathbf{F}_{1}} \underbrace{\begin{bmatrix} 
\mathbf{s}_{1} \\   \mathbf{u}_{1}
\end{bmatrix}}_{\mathbf{x}_{1}} + \underbrace{\begin{bmatrix}
\mathbf{n}_{1}^{1} \\   \mathbf{n}_{1}^{2} \\ \vdots \\  \mathbf{n}_{1}^{R} \\ \tilde{\mathbf{n}}_{1}
\end{bmatrix}}_{\mathbf{n}_{1}} \right).  \label{post_processed_signal} 
\end{align} 
\subsection*{Lower bounding \textbf{Term A}:}
We note that $\mathbf{s}_{1} \rightarrow \mathbf{x}_{1}   \rightarrow  \mathbf{y}_{1} \rightarrow \tilde{\mathbf{y}}_{1}$ forms a Markov chain, thus 
\begin{align}
I(\mathbf{x}_{1}; \mathbf{y}_{1} | \Omega) \geq  I(\mathbf{s}_{1}; \mathbf{y}_{1} | \Omega)  & \geq I(\mathbf{s}_{1}; \tilde{\mathbf{y}}_{1} | \Omega),  \\ 
& = h(\tilde{\mathbf{y}}_{1} | \Omega) - h(\tilde{\mathbf{y}}_{1} | \mathbf{s}_{1}, \Omega). \label{last_step_leakage}
\end{align}
Using Eq. (\ref{post_processed_signal}), we can write $h(\tilde{\mathbf{y}}_{1} | \Omega )$ as follows:
\begin{align}
h(\tilde{\mathbf{y}}_{1} | \Omega)  &= h(\mathbf{\mathbf{\Psi}} (\mathbf{F}_{1} \mathbf{x}_{1} + \mathbf{n}_{1}) ), \\
& = h(\mathbf{F}_{1} \mathbf{x}_{1} + \mathbf{n}_{1})  + \log (\text{det} (\Psi)), \\ 
& = \log(\pi e)^{RT + (n+1)^{N}}   + \sum_{i = 1}^{r(\mathbf{F}_{1})} \log(\lambda_{i} P + \sigma^{2}) + \log (\text{det} (\Psi)), \\ 
& \overset{(a)} = \log(\pi e)^{RT + (n+1)^{N}}   + \sum_{i = 1}^{T} \log(\lambda_{i} P + \sigma^{2}) + \log (\text{det} (\Psi)), \label{last_step_in_entropy_1}
\end{align}
where $\{\lambda_{i}\}_{i=1}^{r(\mathbf{F}_{1})}$ are the singular values of $\mathbf{F}_{1}$.  In (a), we note that  $\mathbf{B}_{1} = \mathbf{\Psi} \mathbf{F}_{1}$ is full rank. Using full rank decomposition Theorem [24], we conclude that  $ \mathbf{F}_{1}$ is also full rank, i.e. $\text{rank}(\mathbf{F}_{1}) = T$.

Now, we write $h(\tilde{\mathbf{y}}_{1} | \mathbf{s}_{1}, \Omega)$ as follows:

\begin{align}
h(\tilde{\mathbf{y}}_{1} | \mathbf{s}_{1}, \Omega)  &= h(\mathbf{\mathbf{\Psi}} (\tilde{\mathbf{F}}_{1} \mathbf{u}_{1} + \mathbf{n}_{1}) ), \\
& = h(\tilde{\mathbf{F}}_{1} \mathbf{u}_{1} + \mathbf{n}_{1})  + \log (\text{det} (\Psi)), \\ 
& = \log( \pi e)^{RT + (n+1)^{N}}   + \sum_{i = 1}^{r(\tilde{\mathbf{F}}_{1})} \log(\lambda_{i}^{'} \ P + \sigma^{2}) + \log (\text{det} (\Psi)), \\  
& \leq  \log( \pi e)^{RT + (n+1)^{N}}   + \sum_{i = 1}^{T_{2}} \log(\lambda_{i}^{'} \ P + \sigma^{2}) + \log (\text{det} (\Psi)), \label{last_step_in_entropy_2}
\end{align}
where,  
\begin{align}
\tilde{\mathbf{F}}_{1}  = \left[ \begin{matrix} \mathbf{H}_{11}^{1} \\ \mathbf{H}_{11}^{2}  \\ \vdots \\ \mathbf{H}_{11}^{R} \\ \tilde{\mathbf{H}}_{11} \mathbf{X} \end{matrix} \right] \mathbf{V}_{1, \mathbf{u}_{1}}, 
\end{align}
where $\tilde{\mathbf{F}}_{1}$ has dimensions of $RT + (n+1)^{N} \times T_{2}$, and $\{\lambda_{i}^{'}\}_{i=1}^{r(\tilde{\mathbf{F}}_{1})}$ are the singular values of  $\tilde{\mathbf{F}}_{1}$.  $\mathbf{V}_{1, \mathbf{u}_{1}}$ has dimensions of $T \times T_{2}$.  Note that,  we can view the \textit{mixing} matrix $\mathbf{V}_{i}$  being composed of two parts i.e., $ \mathbf{V}_{i} =  (\mathbf{V}_{i,\mathbf{s}_{i}} : \mathbf{V}_{i, \mathbf{u}_{i}}  ), \forall i \in 1, 2, \dots, K$ where $\mathbf{V}_{i,\mathbf{s}_{i}} $ corresponding to the information symbol $\mathbf{s}_{i}$ and $\mathbf{V}_{i, \mathbf{u}_{i}}$  corresponding to the artificial noise $\mathbf{u}_{i}$.

From the substitution of (\ref{last_step_in_entropy_1}) and (\ref{last_step_in_entropy_2}) into (\ref{last_step_leakage}), we obtain
\begin{align}
I(\mathbf{x}_{1}; \mathbf{y}_{1} | \Omega) \geq I(\mathbf{s}_{1}; \mathbf{y}_{1} | \Omega)  & \geq I(\mathbf{s}_{1}; \tilde{\mathbf{y}}_{1} | \Omega), \\ 
 & \geq \sum_{i =1}^{T} \log(\lambda_{i} P + \sigma^{2}) -  \sum_{i =1}^{T_{2}} \log(\lambda_{i}^{'} P + \sigma^{2}). \label{information_1}
\end{align}

Before calculating the second term, i.e., \textbf{Term B}. We collectively write the received signal $\mathbf{y}_{j}$ at receiver $j$ over $RT + K (n+1)^{N}$ time slots as follows: 
\begin{align}
\mathbf{y}_{j}= \mathbf{A}_{j} \mathbf{V} \mathbf{q} + \mathbf{n}_{j},  \label{adversary_equation}
\end{align} 

where  $\mathbf{A}_{j}$ is written as follows: 

\begin{align}
 &\mathbf{A}_{j} =  \begin{bmatrix}
\mathbf{C}_{j} \\ \hline  \mathbf{D}_{j}
\end{bmatrix}, \hspace{0.05in} \mathbf{C}_{j} =  \left[ \begin{matrix} \mathbf{H}_{j1}^{1} & \mathbf{H}_{j2}^{1} & \cdots  & \mathbf{H}_{jK}^{1} \\ \mathbf{H}_{j1}^{2} & \mathbf{H}_{j2}^{2} & \cdots  & \mathbf{H}_{jK}^{2} \\ \vdots  & \vdots  & \cdots  & \vdots \\ \mathbf{H}_{j1}^{R} & \mathbf{H}_{j2}^{R} & \cdots  & \mathbf{H}_{jK}^{R} \end{matrix} \right], \nonumber \\ 
&\mathbf{D}_{j} = \text{blkdiag}(\tilde{\mathbf{H}}_{j1} \mathbf{X}, \dots, \tilde{\mathbf{H}}_{jK} \mathbf{X}).
\end{align}
where $\mathbf{A}_{j}$ has dimensions of $(RT + K (n+1)^{N}) \times KT$. The matrix $V$ is written as follows: 
\begin{align}
\mathbf{V} = \text{blkdiag} (\mathbf{V}_{1}, \mathbf{V}_{2}, \dots, \mathbf{V}_{K}),
\end{align}
where $\mathbf{V}$ is the block diagonal matrix with dimensions of $KT \times KT$. Furthermore, we  write
\begin{align}
\mathbf{q} = \left[ \begin{matrix} \mathbf{s}_{1}^{T} & \mathbf{u}_{1}^{T} & \mathbf{s}_{2}^{T} & \mathbf{u}_{2}^{T} & \cdots  & \mathbf{s}_{K}^{T} & \mathbf{u}_{K}^{T} \end{matrix} \right]^{T}
\end{align}
as a column vector of length $KT$, which contains the information symbols and the artificial noises of transmitters $1, \dots, K$.

\subsection*{Upper bounding \textbf{Term B}:}

Now, we  can compute \textbf{Term B}, i.e., $I(\mathbf{x}_{1}; \mathbf{y}_{j} | \Omega)$ as follows: 
\begin{align}
I(\mathbf{x}_{1}; \mathbf{y}_{j} | \Omega) & = h(\mathbf{y}_{j} | \Omega) - h(\mathbf{y}_{j} | \mathbf{x}_{1}, \Omega ), \\ 
& = h(\mathbf{A}_{j} \mathbf{V} \mathbf{q} + \mathbf{n}_{j} ) - h(\tilde{\mathbf{A}}_{j} \tilde{\mathbf{V}} \tilde{\mathbf{q}} + \mathbf{n}_{j} ), \label{leakage_term_information}
\end{align}

\noindent where $\tilde{\mathbf{A}}_{j}$ is a truncated version of $\mathbf{A}_{j}$ with dimensions $(RT+K(n+1)^N) \times (K-1)T$. The matrix $\tilde{\mathbf{V}}$ is written as follows: 
\begin{align}
\tilde{\mathbf{V}} = \text{blkdiag} ( \mathbf{V}_{2}, \dots, \mathbf{V}_{K}),
\end{align}
where $\tilde{\mathbf{V}}$ is the block diagonal matrix with dimensions of $(K-1) T \times (K-1) T$. Furthermore, we  write
\begin{align}
\tilde{\mathbf{q}} = \left[ \begin{matrix}   \mathbf{s}_{2}^{T} & \mathbf{u}_{2}^{T} & \cdots  & \mathbf{s}_{K}^{T} & \mathbf{u}_{K}^{T} \end{matrix} \right]^{T}
\end{align}
as a column vector of length $(K-1) T$, which contains $(K-1) T_{1}$ information symbols and $(K-1) T_{2}$ artificial noises of transmitters.

Using equation (\ref{adversary_equation}), we can write $h(\mathbf{y}_{j} | \Omega)$ as follows:  
\begin{align}
h(\mathbf{y}_{j} | \Omega)   & \overset{(a)} = h(\mathbf{A}_{j} \mathbf{V} \mathbf{q} + \mathbf{n}_{j}) \\
& =  \log( \pi e)^{RT + K (n+1)^{N}}   + \sum_{i = 1}^{r(\mathbf{A}_{j} \mathbf{V})} \log(\Lambda_{i} P + \sigma^{2}), \\
& \overset{(b)}  =  \log( \pi e)^{RT + K (n+1)^{N}}   + \sum_{i = 1}^{r(\mathbf{A}_{j})} \log(\Lambda_{i} P + \sigma^{2}).  \label{sub_1}
\end{align}
In (a), we used Lemma \ref{lemma_one}. $\{\Lambda_{i}\}_{i=1}^{r(\mathbf{A}_{j} \mathbf{V})}$ are the singular values of $\mathbf{A}_{j} \mathbf{V}$. Note that in (b), $\mathbf{V}$ is an invertible matrix with rank $(K-1)T$, therefore $\text{rank} (\mathbf{A}_{j} \mathbf{V}) = \text{rank} (\mathbf{A}_{j})$.

Now, we write $h(\mathbf{y}_{j} | \mathbf{x}_{1}, \Omega) $ as follows: 
\begin{align}
h(\mathbf{y}_{j} | \mathbf{x}_{1}, \Omega)  
& = \log( \pi e)^{RT + K (n+1)^{N}}   + \sum_{i = 1}^{r(\mathbf{A}_{j} \tilde{\mathbf{V}})} \log(\Lambda_{i}^{'} P + \sigma^{2}), \\ 
& \overset{(a)} \geq \log( \pi e)^{RT + K (n+1)^{N}}   + \sum_{i = 1}^{r(\mathbf{A}_{j} \tilde{\tilde{\mathbf{V}}})} \log(\Lambda_{i}^{'} P + \sigma^{2}), \\
& \overset{(b)} = \log( \pi e)^{RT + K (n+1)^{N}}   + \sum_{i = 1}^{r(\mathbf{A}_{j})} \log(\Lambda_{i}^{'} P + \sigma^{2}), \label{sub_2}
\end{align}
where $\{\Lambda_{i}^{'}\}_{i=1}^{r(\mathbf{A}_{j} \tilde{\mathbf{V}})}$ are the singular values of $\mathbf{A}_{j} \tilde{\mathbf{V}}$. In (a), $ \tilde{\tilde{\mathbf{V}}}$ is a truncated version of $\tilde{\mathbf{V}}$ with dimensions of $KT \times (RT+K(n+1)^{N})$, therefore, $r(\mathbf{A}_{j} \tilde{\mathbf{V}}) \geq r(\mathbf{A}_{j} \tilde{\tilde{\mathbf{V}}})  $.   In (b), we used Lemma \ref{lemma_2}, i.e., $\text{rank}(\mathbf{A}_{j}  \tilde{\tilde{\mathbf{V}}}) = \text{rank}(\mathbf{A}_{j}) $.  The multiplication of $\mathbf{A}_{j}$ and $\tilde{\tilde{\mathbf{V}}}$ can be viewed as $RT + K(n+1)^{N}$ linear combinations of the $KT$ rows of matrix $ \tilde{\tilde{\mathbf{V}}}$, whose elements are generated independently of $\mathbf{A}_{j}$ from a continuous distribution. In Appendix XI of \cite{seif2018arxiv}, we show that  multiplying  $\mathbf{A}_{j}$ with a non-square random matrix $ \tilde{\tilde{\mathbf{V}}}$ does not reduce the rank of matrix $\mathbf{A}_{j}$, almost surely. Hence, from the above argument,  in order to ensure that $\text{rank}(\mathbf{A}_{j} \tilde{\tilde{\mathbf{V}}}) = \text{rank}(\mathbf{A}_{j})$, we must pick $RT + (K-1) (n+1)^{N} \leq (K-1) T_{2}$, which gives the reasoning behind the choice of the parameter $T_2$.

From the substitution of (\ref{sub_1}) and (\ref{sub_2}) into (\ref{leakage_term_information}), we obtain 
\begin{align}
I (\mathbf{x}_{1}; \mathbf{y}_{j} | \Omega)  \leq  \sum_{i = 1}^{r(\mathbf{A}_{j})} \log(\Lambda_{i} P + \sigma^{2}) - \sum_{i = 1}^{r(\mathbf{A}_{j})} \log(\Lambda_{i}^{'} P + \sigma^{2}),  \label{information_2}
\end{align}
Combining (\ref{information_1}) and (\ref{information_2}), we have 
\begin{align}
& R_{1}  \geq \frac{\sum_{i=1}^{T} \log(\lambda_{i} P + \sigma^{2} ) - \sum_{i=1}^{T_{2}} \log(\lambda_{i}^{'} P + \sigma^{2}) - \left( \sum_{i=1}^{r(\mathbf{A}_{j})} \log(\Lambda_{i} P + \sigma^{2}) -  \sum_{i=1}^{r(\mathbf{A}_{j})} \log(\Lambda_{i}^{'} P + \sigma^{2})  \right) }{ RT + K (n+1)^{N}}, \\
& \geq  \underbrace{\frac{T \log(\lambda_{\min} P + \sigma^{2}) - T_{2} \log(\lambda_{\max}^{'} P + \sigma^{2}) }{RT + K (n+1)^{N}}}_{\text{Term 1}}  
 - \underbrace{\frac{ \left( r(\mathbf{A}_{j}) \log(\Lambda_{\max} P + \sigma^{2}) - r(\mathbf{A}_{j})  \log(\Lambda_{\min}^{'} P + \sigma^{2})  \right)}{RT + K (n+1)^{N}}}_{\text{Term 2}}, \label{rate_expression}
\end{align}
where $\lambda_{\min} = \min_{i} \{\lambda_{i} \}_{i=1}^{r(\mathbf{F}_{1})}$, $\lambda_{\max}^{'} = \max_{i} \{\lambda_{i}^{'}\}_{i=1}^{r(\mathbf{\tilde{F}}_{1})}$, $\Lambda_{\min}^{'} = \min_{i} \{ \Lambda_{i}^{'} \}_{i=1}^{r(\mathbf{A_{j} \tilde{V}})}$ and $\Lambda_{\max} = \max_{i} \{ \Lambda_{i} \}_{i=1}^{r(\mathbf{A_{j} V})}$.

We now simplify the two terms as follows: 
\begin{align}
& \text{Term $1$} = \frac{T \log(\lambda_{\min} P + \sigma^{2}) - T_{2} \log(\lambda_{\max}^{'} P + \sigma^{2})}{RT + K (n+1)^{N}},  \\ 
& = \frac{ \left( Rn^{N} + (n+1)^{N} \right) \log(\lambda_{\min} P + \sigma^{2})  -  \bigg{\lceil} \frac{R(R n^{N} + (n+1)^{N}) + (K-1) (n+1)^{N} }{K-1} \bigg{\rceil}\log(\lambda_{\max}^{'} P + \sigma^{2})  }{R (R n^{N} + (n+1)^{N}) + K (n+1)^{N}},  \\ 
&  \overset{(a)} \geq \frac{ \left( Rn^{N} + (n+1)^{N} \right) \log(\lambda_{\min} P + \sigma^{2})  -  \left( \frac{R(R n^{N} + (n+1)^{N}) + (K-1) (n+1)^{N} }{K-1} + 1 \right)\log(\lambda_{\max}^{'} P + \sigma^{2})  }{R (R n^{N} + (n+1)^{N}) + K (n+1)^{N}},  \label{term_one_rate}
\end{align}
where in (a), we used the property that $\lceil x \rceil < x+1$.

\noindent Also, 
\begin{align}
\text{Term $2$} & =  \frac{r(\mathbf{A}_{j})   \left( \log(\Lambda_{\max} P + \sigma^{2}) -  \log(\Lambda_{\min}^{'} P + \sigma^{2})  \right)}{RT + K (n+1)^{N}}, \\ 
& = \frac{r(\mathbf{A}_{j})   \left( \log(\Lambda_{\max} P + \sigma^{2}) -  \log(\Lambda_{\min}^{'} P + \sigma^{2})  \right)}{R(Rn^{N} + (n+1)^{N}) + K (n+1)^{N}}, \\ 
& \leq \frac{\left( R(Rn^{N} + (n+1)^{N}) + K (n+1)^{N} \right)  \left( \log(\Lambda_{\max} P + \sigma^{2}) -  \log(\Lambda_{\min}^{'} P + \sigma^{2})  \right)}{R(Rn^{N} + (n+1)^{N}) + K (n+1)^{N}}, \\ 
& = \log(\Lambda_{\max} P + \sigma^{2}) -  \log(\Lambda_{\min}^{'} P + \sigma^{2}). \label{term_two_rate}
\end{align}
Combining (\ref{term_one_rate}) and (\ref{term_two_rate}) in (\ref{rate_expression}) and taking the limit $n \rightarrow \infty$, we get the following: 
\begin{align}
\lim_{n \rightarrow \infty} R_{1} & =  \frac{(R +1 ) \log(\lambda_{\min} P + \sigma^{2})   - \left( \frac{R (R +1)}{K-1}  + 1\right)  \log(\lambda_{\max}^{'} P + \sigma^{2}) }{ R (R+1) + K}  \nonumber \\
& \hspace{0.4in} - \log(\Lambda_{\max} P + \sigma^{2}) +  \log(\Lambda_{\min}^{'} P + \sigma^{2}).
\end{align}
Dividing $R_{1}$ by $\log(P)$ and letting $P \rightarrow \infty$, we get 
\begin{align}
d_{1} =  \lim_{P \rightarrow \infty} \frac{R_{1}}{\log (P)} & =  \frac{(R+1) - \left( \frac{R^{2} + R}{K-1} +1\right)}{ R (R+1) + K}, \nonumber \\ 
& = \frac{R (K-R-2)}{(K-1) \times  \left[ R(R+1) + K\right]}.
\end{align}
Therefore,  the achievable secure sum degrees of freedom $(\text{SDoF}_{\text{IC-CM}}^{\text{ach.}}) $ is obtained as
\begin{align}
 \text{SDoF}_{\text{IC-CM}}^{\text{ach.}} =  \frac{K R (K-R-2)}{ (K-1) \times \left[R (R+1)+K\right]}.
\end{align}
Hence, this completes the proof of Theorem 1.

\section{Proof of Theorem \ref{theorem::2} } \label{appendix_e}

\indent We follow a similar achievability scheme presented in Section \ref{seciton_proof_theorem}, however, the main differences are the number of information symbols, the artificial noises used for transmission and the number of rounds in the first phase of the scheme. The goal of each transmitter is to securely send $T_{1} = Rn^{N} + (n+1)^{N} - \lceil \frac{RT + K (n+1)^{N}}{K}\rceil$ information symbols to its corresponding receiver and keeping all messages secure against the external eavesdropper. 

\indent The total number of equations seen at the eavesdropper is $RT + K(n+1)^{N}$ of  $\{\mathbf{s}_{i}\}_{i=1}^{K}$,  $\{\mathbf{u}_{i}\}_{i=1}^{K}$. Hence, in order to keep the unintended information symbols of $K$ transmitters at this receiver secure, we require that the number of these equations must be at most equal to the total number of the artificial noise dimensions of the $K$ transmitters, i.e.,
\begin{align}
RT + K (n+1)^{N} \leq K T_{2}.
\end{align}
Therefore, we choose $T_{2}$ as
\begin{align}
T_{2} =  \bigg{\lceil} \frac{RT + K (n+1)^{N} }{K} \bigg{\rceil}.  \label{mind}
\end{align}
Since $T = T_{1} + T_{2}$, so we can get $T_{1}$  as follows:
\begin{align}
 T_{1} = Rn^{N} + (n+1)^{N} -  \bigg{\lceil} \frac{RT +  K (n+1)^{N}}{K } \bigg{\rceil}, \hspace{0.1in} T = Rn^{N} + (n+1)^{N}.
\end{align}
To this end, this scheme leads to the following achievable SDoF:
\begin{align}
 \text{SDoF}_{\text{IC-EE}}^{\text{ach.}}
& \overset{(b)} = \frac{R (K-R-1)}{ R (R+1)+K }. \label{approximated_EE}
\end{align}
Since the achieved SDoF in (\ref{approximated_EE}) is a concave function of $R$. Hence, getting the optimal $R^{*}$ is obtained by equating the first derivative of the function with zero. Therefore, the optimal $R^{*}$ is 
\begin{align}
R^{*} & = \lfloor \sqrt{K} \rfloor -1, \\ 
& > \sqrt{K} - 2. 
\end{align}
Now we approximate the obtained SDoF as follows: 
\begin{align}
 \text{SDoF}_{\text{IC-EE}}^{\text{ach.}}& = \frac{(\sqrt{K} - 2 ) (K - \sqrt{K} + 1)}{ (\sqrt{K} - 2) (\sqrt{K} - 1) + K}, \\
& \overset{(a)}  = \frac{K \sqrt{K} - 3 K + 3 \sqrt{K} -2}{2K - 3 \sqrt{K} + 2 }, \\
& >  \frac{K \sqrt{K} - 3 K + 3 \sqrt{K} -2}{2K}, \label{equation_ee} \\ 
& \overset{(b)}  = \frac{\sqrt{K} - 3 }{2} + \frac{3\sqrt{K} -2 }{2K}, \\
&> \frac{1}{2} (\sqrt{K} - 3)^{+}, \label{last_ee}
\end{align}
where in (a), the term $-3\sqrt{K} + 2 $ in the denominator is negative, $\forall K \geq  1$,  so neglecting this term gives us  (\ref{equation_ee}). In step (b), since the term $3 \sqrt{K} -2$ is positive, hence omitting this term gives (\ref{last_ee}). 

\subsection{Secrecy Rate and SDoF Calculation} \label{rate_analysis_EE}

For a transmission of block length $B = RT + K(n+1)^{N}$, the  achievable secure rate $R_{i}, i = 1, 2, \dots, K$ is defined as 
\begin{align}
R_{i} = \frac{ I(\mathbf{x}_{i}; \mathbf{y}_{i} | \Omega) -  I(\mathbf{x}_{i}; \mathbf{z}  | \Omega  )} {R T + K (n+1)^{N}}, \hspace{0.05in} i = 1,2, \dots, K,  \label{korner_rate}
\end{align} 
where $I(\mathbf{x}_{i}; \mathbf{y}_{i} | \Omega)$ is the mutual information between the transmitted symbols vector $\mathbf{x}_{i}$ and $\mathbf{y}_{i}$,  the received composite signal vector at the intended receiver $i$, given the knowledge of the channel coefficients.  $I(\mathbf{x}_{i}; \mathbf{z} | \Omega)$ is the mutual information between $\mathbf{x}_{i}$ and $\mathbf{z}$, the received composite signal vector at the external eavesdropper. Note that $\mathbf{z}$ is collectively written as 
\begin{align}
\mathbf{z} = \mathbf{A}_{z} \mathbf{V}  \mathbf{q}_{z} + \mathbf{n}_{z},
\end{align}
where,

\begin{align}
 &\mathbf{A}_{z} =  \begin{bmatrix}
\mathbf{C}_{z} \\ \hline  \mathbf{D}_{z}
\end{bmatrix}, \hspace{0.05in} \mathbf{C}_{z} =  \left[ \begin{matrix} \mathbf{G}_{1}^{1} & \mathbf{G}_{2}^{1} & \cdots  & \mathbf{G}_{K}^{1} \\ \mathbf{G}_{1}^{2} & \mathbf{G}_{2}^{2} & \cdots  & \mathbf{G}_{K}^{2} \\ \vdots  & \vdots  & \cdots  & \vdots \\ \mathbf{G}_{1}^{R} & \mathbf{G}_{2}^{R} & \cdots  & \mathbf{G}_{K}^{R} \end{matrix} \right], \nonumber \\ 
&\mathbf{D}_{z} = \text{blkdiag}(\tilde{\mathbf{G}}_{1} \mathbf{X}, \dots, \tilde{\mathbf{G}}_{K} \mathbf{X}).
\end{align}

\noindent where  $\mathbf{A}_{z}$ has dimensions of $(RT + K (n+1)^{N}) \times KT = K T_{2} \times KT$.  Each $\mathbf{G}$ is a matrix represents the channel gains between each transmitter and the external eavesdropper.

The analysis of the achievable secure rate and SDoF follows similar steps as those in subsection \ref{rate_analysis_CM}. This completes the proof of Theorem \ref{theorem::2}.  


\section{Proof of Theorem \ref{theorem::3} } \label{appendix_f} 

We follow the same transmission scheme presented in Section \ref{seciton_proof_theorem}. The goal of each transmitter is to securely send $T_{1} = Rn^{N} + (n+1)^{N} - \lceil \frac{RT +( K-1) (n+1)^{N}}{K-1}\rceil$ information symbols to its corresponding receiver and keeping all messages secure against the external eavesdropper and the unintended receivers.

\indent We have two secrecy constraints must be satisfied, i.e., 
\begin{align}
RT + (K-1) (n+1)^{N} & \leq (K-1) T_{2}, \hspace{0.05in} \text{(confidential messages)}, \\
RT + K (n+1)^{N} & \leq K T_{2}, \hspace{0.05in} \text{(eavesdropper)}, \label{equation_needed}
\end{align}
Equation (\ref{equation_needed}) can be re-written as
\begin{align}
RT + (K-1) (n+1)^{N}  + (n+1)^{N} & \leq (K-1) T_{2} + T_{2}.
\end{align}
So if we pick $T_{2}$ as 
\begin{align}
T_{2} =   \bigg{\lceil} \frac{RT + (K-1) (n+1)^{N} }{K-1} \bigg{\rceil}
\end{align}
we need to check that $T_{2} \geq (n+1)^{N}$. $T_{2}$ can be written as 
\begin{align}
T_{2} & = \bigg{ \lceil}  \frac{R (R n^{N} + (n+1)^{N}) + (K-1) (n+1)^{N}}{K-1} \bigg{ \rceil}, \\
& > \frac{R (R n^{N} + (n+1)^{N}) + (K-1) (n+1)^{N}}{K-1}  - 1, \\
& = \frac{R^{2} n ^{N} + R (n+1)^{N} + (K-1) (n+1)^{N}}{K-1} -1, \\ 
& \overset{(a)} = \frac{R^{2}}{(K-1)} n^{N} + \frac{R}{K-1} (n+1)^{N} + (n+1)^{N} - 1 > (n+1)^{N}, 
\end{align}
where in (a), the first two terms are positive, hence, $T_{2}$ is strictly greater than $(n+1)^{N}$. To this end, we conclude that the two secrecy constraints are satisfied. Hence, we achieve the same SDoF of Theorem \ref{theorem::1}, i.e., 
\begin{align}
 \text{SDoF}_{\text{IC-CM-EE}}^{\text{ach.}} & =  \frac{K R (K-R-2)}{ (K-1) \times \left[R (R+1)+K\right]}  > \frac{1}{2} (\sqrt{K} - 6)^{+}.
\end{align}

\subsection{Secrecy Rate and SDoF Calculation} \label{rate_analysis_EE_CM}

For a transmission of block length $B = RT + K(n+1)^{N}$, the  achievable secure rate $R_{i}, i = 1, 2, \dots, K$ is defined as 
\begin{align}
R_{i} = \frac{ I(\mathbf{x}_{i}; \mathbf{y}_{i} | \Omega) -  \max \left[ \max_{j \neq i} I(\mathbf{x}_{i}; \mathbf{y}_{j}  | \Omega  ), I(\mathbf{s}_{i}; \mathbf{z}  | \Omega  )  \right]} {R T + K (n+1)^{N}}, \hspace{0.05in} i = 1,2, \dots, K. \label{korner_rate}
\end{align}

The analysis of the achievable secure rate and SDoF follows similar steps as those in subsection \ref{rate_analysis_CM}. This completes the proof of Theorem \ref{theorem::3}.


\section{Conclusion} \label{conclusion}
In this paper, we studied the $K$-user  interference channel with three secrecy constrained channel models and delayed CSIT: we showed  that for the $K$-user interference channel with confidential messages, the sum secure degrees of freedom (SDoF)  is  at least $\frac{1}{2} (\sqrt{K} -6)$, and scales with  {  square root of the number of users.} Also, we showed that for the $K$-user interference channel  with an external eavesdropper,  $\frac{1}{2} (\sqrt{K} -3) $ SDoF is achievable. For the $K$-user interference channel with confidential messages and an external eavesdropper, we showed that $\frac{1}{2} (\sqrt{K} -6) $ is achievable. To achieve these results,  we have proposed novel  secure retrospective interference alignment schemes which satisfy both  secrecy and decodability at receivers.  To the best of our knowledge,  this is the first result showing scaling of SDoF for the interference channel with secrecy constraints and delayed CSIT.  An interesting open problem is to investigate the optimality of these schemes, and finding upper bounds on SDoF with delayed CSIT for these channel models.



\section*{Appendices}

\section{ Proof of Lemma \ref{concavity_proof} and Corollary \ref{cor1_proof}}  \label{section_a}
\noindent By taking the first derivative of (\ref{approximated}) with respect to the number of rounds $R$, we get
\begin{align}
\frac{\partial}{\partial R}  \hspace{0.05in} \text{SDoF}_{\text{IC-CM}}^{\text{ach.}} (K,R)) = \frac{K(K^{2} - K (R^{2} + 2R +2) + R^{2})}{(K-1) (K+R^{2} + R)^{2}}.
\end{align}
For $R < R^{*}$, the function   $\text{SDoF}_{\text{IC-CM}}^{\text{ach.}} (K,R)$ strictly increases and for $R > R^{*}$ the function strictly decreases, where  $R^{*}$ is given by
\begin{align}
R^{*}   =  \frac{-K+K \times \sqrt{1 +  \frac{(K-1)(K-2)}{K}}}{K-1} . 
\end{align}
Alternatively, 
\begin{align}
\frac{\partial}{\partial R}  \hspace{0.05in} \text{SDoF}_{\text{IC-CM}}^{\text{ach.}} (K,R)  &> 0, \forall R < R^{*}, \\
\frac{\partial}{\partial R}  \hspace{0.05in} \text{SDoF}_{\text{IC-CM}}^{\text{ach.}} (K,R) &< 0, \forall R > R^{*}.
\end{align}
\begin{figure}[h]
  \centering
\includegraphics[scale=0.45]{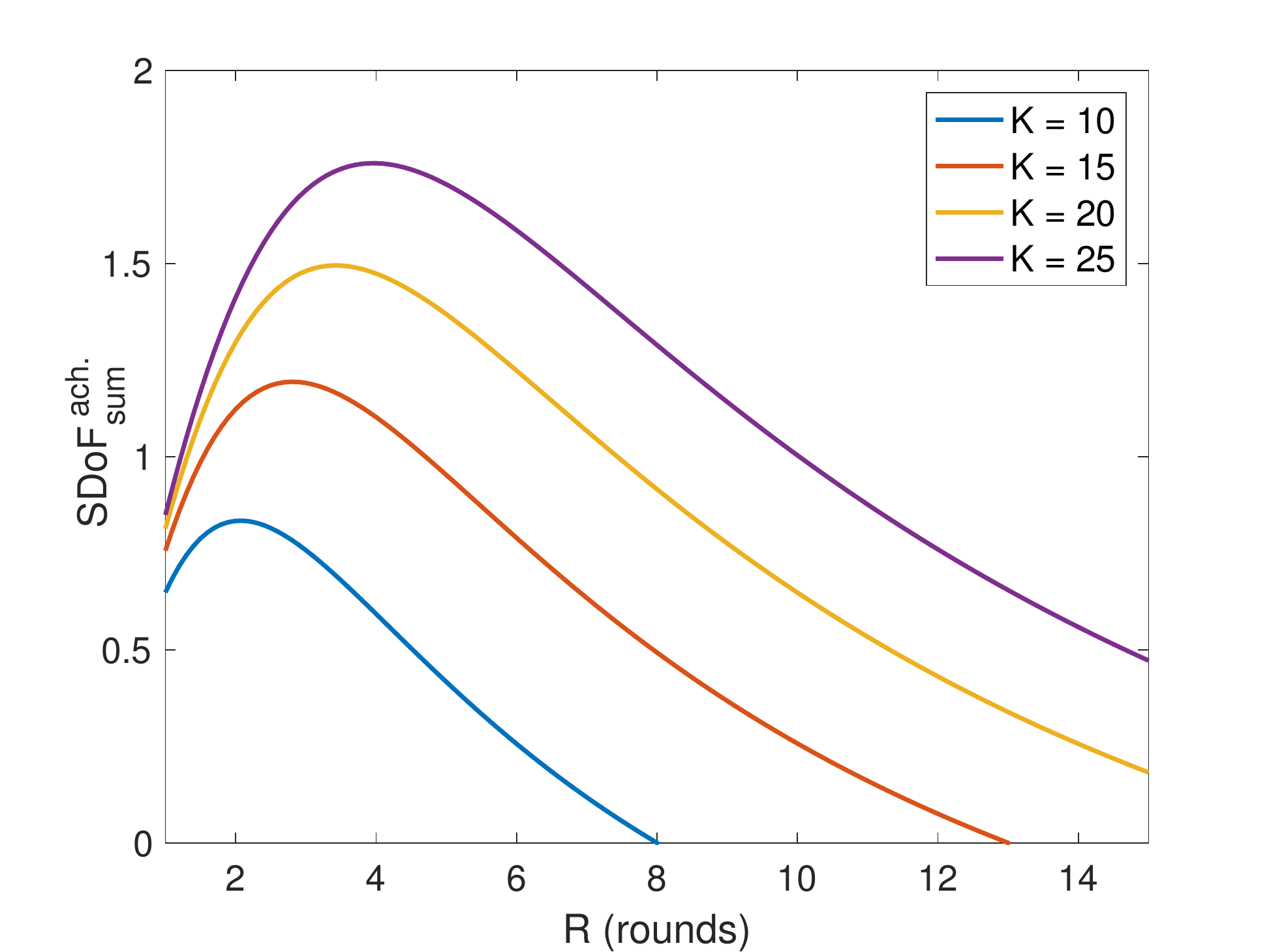}
\caption{Plot for the achievable sum SDoF as a function of the number of rounds $R$ for different number of user $K$. }
\label{fig::concavity}
\end{figure}
\noindent Fig. \ref{fig::concavity} shows the behavior of the achievable sum SDoF as a function of the number of rounds $R$. 
The optimal value of $R$ can be obtained by equating the first derivative of $\text{SDoF}_{\text{sum}}$ to zero as follows:
\begin{align}
\frac{\partial }{\partial R} \text{SDoF}_{\text{IC-CM}}^{\text{ach.}} (K,R) K,R) & = \frac{\partial }{\partial R} \frac{K R (K-R-2)}{ (K-1) \times \left[ R (R+1)+K\right]} = 0,\\
& \varpropto \frac{\partial}{\partial R}  \frac{ R (K-R-2)}{  R (R+1)+K} = 0. \label{differentiation}
\end{align}
After differentiating equation (\ref{differentiation}), we will get the following:
\begin{align}
R^{2} (K-1) + 2KR - K(K-2) = 0.
\end{align}
The solution of the previous equation is
\begin{align}
R^{*}  & \overset{(a)} =  \bigg{\lfloor} \frac{-K+K \times \sqrt{1 +  \frac{(K-1)(K-2)}{K}}}{K-1} \bigg{\rfloor}, \\ 
&>  \bigg{\lfloor} \frac{-K+K \times \sqrt{1 +  \frac{(K-1)(K-2)}{K}}}{K} \bigg{\rfloor}, \\
&=  \bigg{\lfloor}  -1+ \sqrt{1 +  \frac{(K-1)(K-2)}{K}} \bigg{\rfloor},\\
& \overset{(b)}  >    \sqrt{\frac{(K-1)(K-2)}{K}} -2,\\
& =    \sqrt{\frac{K^{2} - 3K +2}{K}} -2 >    \sqrt{\frac{K^{2} - 3K}{K}} -2  =  \sqrt{K -3} -2,\\
& \overset{(c)} >   \sqrt{K} -3 -2 =  \sqrt{K} -5  = R_{\text{lb}},
\end{align}
where in (a), since  $R^{*} \in \mathds{N}$,  we apply the floor rounding operator on the obtained value of $R$. In (b), we used the property of the floor operator, i.e., $\lfloor x \rfloor > x -1$.  In (c), the term $\sqrt{K-3}$ is greater than $\sqrt{K} -3, \forall K \geq 1$.
\section{ Proof of Linear Independence in (\ref{important_equation}) }\label{linear_indpendence_proof}
\noindent In this Appendix, we show that by the end of the transmission scheme, each receiver gets $T = R  n^{N} + (n+1)^{N}$ linear independent equations of the desired signals (i.e., the information symbols and the artificial noises). Then, we need to show that the following matrix 
\begin{align}
\mathbf{B}_{k} = \left[\mathbf{X}^{H}, (\mathbf{W}^{H} \mathbf{H}_{kk}^{1})^{H}, \dots, (\mathbf{W}^{H}\mathbf{H}_{kk}^{R})^{H}\right]^{H}
\end{align}
is full rank. Since  $\mathbf{B}_{k}$ is a square matrix, then it is sufficient to show that $\text{det}(\mathbf{B}_{k}) \neq 0, \forall k = 1, 2, \dots, K$.
Without loss of generality, we consider receiver $1$ which has the following matrix
\begin{align}
\mathbf{B}_{1} = \left[\mathbf{X}^{H}, (\mathbf{W}^{H} \mathbf{H}_{11}^{1})^{H}, \dots, (\mathbf{W}^{H}\mathbf{H}_{11}^{R})^{H}\right]^{H}.
\end{align}
Since $\text{det}(\mathbf{B}_{1}) = \text{det}(\mathbf{B}_{1}^{H}) $, we will instead show that $ \text{det}(\mathbf{B}_{1}^{H}) \neq 0$, which is given as follows:
\begin{align}
\mathbf{B}_{1}^{H} = \left[\mathbf{X}^{H},  (\mathbf{H}_{11}^{1})^{H} \mathbf{W} , \dots, (\mathbf{H}_{11}^{R})^{H} \mathbf{W} \right].
\end{align}
Note that $\mathbf{W}$ and $\mathbf{X}$ are
 function of the diagonal entries of the channels  $\{(\mathbf{H}_{kj}^{r})^{H}\}_{k \neq j}, \\
  \forall k, j = 1, \dots, K $ and $ r = 1, 2, \dots, R $. More specifically, the  entries of $(\mathbf{H}_{kj}^{r})^{H}$ are $h_{kj}^{r}(t), \forall t = 1, 2, \dots, T$.  $\mathbf{B}_{1}$ depends on $\{{(\mathbf{H}_{kj}^{r})^{H}\}_{k \neq j}}$ plus $(\mathbf{H}_{11}^{r})^{H}$ whose elements are $h_{11}^{r}(t), \forall t=1, \dots, T$ and, $\forall r = 1, \dots, R$. For notation convenience, let us denote these channel coefficients as $\mu_{mt}, \forall t=1,2, \dots, T$ and $\forall m =1, 2, \dots, N+R$.  The elements of  $\mathbf{B}_{1}^{H}$ are written as a monomial function of the random variables $\mu_{it}, \forall  i=1, \dots, N+R $ and $\forall t = 1, 2, \dots, T $ as follows:
\begin{align}
B_{1}^{H}(t, p) = \prod_{i=1}^{N+R} (\mu_{it})^{n_{i}(p)} ,
\end{align}
where $n_{i}(p) \in \mathds{Z}_{+}$ is the exponent of the random variable $\mu_{it}$. Note that for two different columns $p_{1}$ and $p_{2}$, $(n_{1}(p_{1}), n_{2}(p_{1}), \dots, n_{N+R}(p_{1})) \neq (n_{1}(p_{2}), n_{2}(p_{2}), \dots, n_{N+R}(p_{2}))$. More specifically, the structure of  $\mathbf{X}^{H}$ is as follows:
\begin{align}
\mathbf{X}^{H} = \left[ \begin{matrix} 1 & \mu_{11} & \mu_{21} & \cdots  & \mu_{11}^{n} \mu_{21}^{n} \dots \mu_{N1}^{n} \\ 1 & \mu_{12} & \mu_{22} & \cdots  & \mu_{12}^{n} \mu_{22}^{n} \dots \mu_{N2}^{n} \\ \vdots  & \vdots  & \vdots  & \ddots  & \vdots  \\ 1 & \mu_{1T}^{n} & \mu_{2T}^{n} & \cdots  & \mu_{1T}^{n} \mu_{2T}^{n} \dots \mu_{NT}^{n} \end{matrix} \right], 
\end{align}
and for $(\mathbf{H}_{11}^{r})^{H}\mathbf{W}$ as 
\begin{align}
(\mathbf{H}_{11}^{r})^{H}\mathbf{W} = \left[ \begin{matrix} \kappa_{1}^{r}
 & \kappa_{1}^{r}
\mu_{11} & \kappa_{1}^{r}
\mu_{21} & \cdots  & \kappa_{1}^{r}
\chi_{1}^{n-1} \\ \kappa_{2}^{r}
 & \kappa_{2}^{r}
\mu_{12} & \kappa_{2}^{r}
\mu_{22} & \cdots  & \kappa_{2}^{r}
\chi_{2}^{n-1}\\ \vdots  & \vdots  & \vdots  & \ddots  & \vdots  \\ \kappa_{T}^{r}
 & \kappa_{T}^{r}
\mu_{1T}^{n} &\kappa_{T}^{r}
 \mu_{2T}^{n} & \cdots  & \kappa_{T}^{r}
\chi_{T}^{n-1} \end{matrix} \right] ,
\end{align}
 where  $\kappa _{ t }^{ r } = \mu_{N+r, t}, \forall t= 1, 2, \dots, T, \forall r= 1, 2, \dots, R$ and  $\chi_{t}^{n-1} = \mu_{1t}^{n} \mu_{2t}^{n} \dots \mu_{Nt}^{n}$. The full matrix $\mathbf{B}_{1}^{H}$ is written as follows:
 
\begin{equation}\label{full_matrix}
\mathbf{B}_{1}^{H} = \left[ \begin{matrix} 1 & \cdots  & \chi _{ 1 }^{ n } & \kappa _{ 1 }^{ 1 } & \cdots  & \kappa _{ 1 }^{ 1 }\chi _{ 1 }^{ n-1 } & \cdots  & \kappa _{ 1 }^{ R } & \cdots  & \kappa _{ 1 }^{ R }\chi _{ 1 }^{ n-1 } \\ 1 & \cdots  & \chi _{ 2 }^{ n } & \kappa _{ 2 }^{ 1 } & \cdots  & \kappa _{ 2 }^{ 1 }\chi _{ 2 }^{ n-1 } & \cdots  & \kappa _{ 2 }^{ R } & \cdots  & \kappa _{ 2 }^{ R }\chi _{ 2 }^{ n-1 } \\ \vdots  & \ddots  & \vdots  & \vdots  & \ddots  & \vdots  & \ddots  & \vdots  & \ddots  & \vdots  \\ 1 & \cdots  & \chi _{ T }^{ n } & \kappa _{ T }^{ 1 } & \cdots  & \kappa _{ T }^{ 1 }\chi _{ T }^{ n-1 } & \cdots  & \kappa _{ T }^{ R } & \cdots  & \kappa _{ T }^{ R }\chi _{ T }^{ n-1 } \end{matrix} \right]  
\end{equation}
 
 \noindent The determinant of matrix $\mathbf{B}_{1}^{H}$ can be written as follows:
\begin{align}
\text{det}(\mathbf{B}_{1}^{H})  = a_{1,1} {C}_{1,1} +  a_{1, 2} {C}_{1,2} + \cdots  + a_{1, T} {C}_{1,T}, \label{determinant}
\end{align}
where $\mathbf{C}_{1,j}$ is the cofactor matrix corresponding after removing the  $1$st row and the $j$th column with coefficient $a_{1,j}$. Now we will show that $\mathbf{B}_{1}^{H}$ is full rank by contradiction. The zero determinant assumption implies one of the following two events: 
\begin{enumerate}
\item $\mu_{m1}, m \in \{1,2, \dots, N+R\}$ takes a value equal to one of the roots of the polynomial equation. 
\item All the cofactors of the polynomials are zero.
\end{enumerate}

\noindent For the first event, none of the cofactors depends on the random variables $\mu_{m1}, m \in \{1, 2, \dots, N+R\}$.  Note that $\mu_{m1}, m \in \{1, 2, \dots, N+R\}$ are drawn from a continuous distribution, then the probability of these random variables that take finitely many values as a solution for the polynomial is zero almost surely.  Therefore,  the second event happens with probability greater than zero, which implies 
\begin{align}
{C}_{1, p}=0, \forall p \in \{1, \dots, T\}.
\end{align}
Then  ${C}_{1, T} = 0$ with probability higher than zero. ${C}_{1, T} = 0$  implies that the determinant of the matrix obtained by stripping off the first row and last column of  $\mathbf{B}_{1}^{H}$ is equal to zero with non-zero probability. Repeating the process of stripping off each row and column, it will end up with $1 \times 1$ matrix with value one which contradicts the assumption that ${C}_{1, T} = 0$. It is worth noting that stripping off the rows and columns procedure preserves the structure of the matrix which means that the cofactors do not depend on the coefficients. To conclude, the determinant of $\mathbf{B}_{1}^{H}$ does not equal zero almost surely, which implies that the desired symbols are decoded successfully at the receiver side with probability one.

 
\section{ Proof of Lemma \ref{lemma_one}  } \label{appendix_c}
\noindent Note that $\mathbf{A} \mathbf{X} + \mathbf{N}$ is a jointly complex Gaussian vector with zero mean and covariance $P \mathbf{A} \mathbf{A}^{H} + \sigma^{2} \mathbf{I}$. From \cite{cover2012elements}, $h(\mathbf{A} \mathbf{X} + \mathbf{N}) $ is written as
\begin{align}
h(\mathbf{A} \mathbf{X} + \mathbf{N}) =  \log(\pi e)^{M} \text{det} \left( P \mathbf{A} \mathbf{A}^{H} + \sigma^{2} \mathbf{I}_{M} \right) .
\end{align} 
It is worth noting  that  $ \mathbf{A} \mathbf{A}^{H} $ is positive semi-definite, with eigenvalue decomposition $\mathbf{Q} \mathbf{D} \mathbf{Q}^{H}$, where $\mathbf{D}$ is a diagonal matrix with  $r$ non-zero eigenvalues $\lambda_{1}, \lambda_{2}, \dots, \lambda_{r}$ where $r =\text{rank}(\mathbf{A} \mathbf{A}^{H}) = \text{rank}(\mathbf{A})$. Then, 
\begin{align}
h(\mathbf{A} \mathbf{X} + \mathbf{N}) &= \log(\pi e)^{M} \text{det} \left( P\mathbf{Q} \mathbf{D} \mathbf{Q}^{H} + \sigma^{2} \mathbf{I}_{M} \right).
\end{align} 
Next, we use the Sylvester's identity for determinants, i.e., $\text{det}(\mathbf{I} + \mathbf{A} \mathbf{B}) =  \text{det}(\mathbf{I} + \mathbf{B} \mathbf{A}) $. Using this identity, we have 
\begin{align}
h(\mathbf{A} \mathbf{X} + \mathbf{N}) = \log(\pi e)^{M} \text{det} \left( P\mathbf{D} \mathbf{Q}^{H} \mathbf{Q}  + \sigma^{2} \mathbf{I}_{M} \right) .
\end{align}
Since $\mathbf{Q}$ is a unitary matrix, (i.e., $\mathbf{Q}^{H} \mathbf{Q} = \mathbf{Q} \mathbf{Q}^{H}  = \mathbf{I} $), we have the following: 
 \begin{align}
h(\mathbf{A} \mathbf{X} + \mathbf{N})  & = \log(\pi e)^{M} \prod_{i=1}^{r} \left( \lambda_{i} P +\sigma^{2} \right), \\
&= \log(\pi e)^{M}+ \sum_{i=1}^{r} \log \left( \lambda_{i} P +\sigma^{2} \right).
\end{align}
This completes the proof of Lemma \ref{lemma_one}.

\section{Proof of Lemma \ref{lemma_2} } \label{appendix_d}

Let us consider two random matrices $\mathbf{A}_{M \times N}$ and $\mathbf{B}_{N \times M}$ where $M \leq N$. The elements of matrix $\mathbf{B}$ are drawn independently from continuous  distribution.  Therefore, $\mathbf{B}$ is full rank (i.e., $\text{rank}(\mathbf{B}) = M$) almost surely. $\mathbf{A}$ has arbitrary structure with $\text{rank}(\mathbf{A})= r(\mathbf{A}) \leq M$. We want to show that,
\begin{align}
\text{rank}(\mathbf{A} \mathbf{B}) =\text{rank}(\mathbf{A}) = r(\mathbf{A}),
\end{align} 
which means that under the previous assumptions, the rank of the matrix product $\mathbf{AB} $ has the same rank of matrix $\mathbf{A}$ almost surely. Our proof steps are similar in spirit of \cite{nafea2013many}. Let us write the matrix $\mathbf{A}$ and $\mathbf{B}$ in terms of their column vectors as follows: 
\begin{align}
\mathbf{A} &=  \left[ \begin{matrix} \mathbf{a}_{1} & \mathbf{a}_{2} & \dots & \mathbf{a}_{N} \end{matrix}  \right], \label{matrix_A}  \\
\mathbf{B} &= \left[ \begin{matrix} \mathbf{b}_{1} & \mathbf{b}_{2} & \dots & \mathbf{b}_{M} \end{matrix}  \right],
\end{align} 
where $\mathbf{a}_{1}, \mathbf{a}_{2}, \dots, \mathbf{a}_{N} $ are the $N$ column vectors of $\mathbf{A}$, each of length $M$. $\mathbf{b}_{1}, \mathbf{b}_{2}, \dots, \mathbf{b}_{M} $ are the $M$  column vectors of $\mathbf{B}$, each of length $N$. Let $b_{j,i}$ denote the entry of $\mathbf{B}$ in the $j$th row and $i$th column. Now we write the matrix $\mathbf{AB}$ as
\begin{align}
\mathbf{AB} =  \mathbf{C} = \left[ \begin{matrix} \mathbf{c}_{1} & \mathbf{c}_{2} & \dots & \mathbf{c}_{M} \end{matrix}  \right],
\end{align}
where $\mathbf{c}_{i}$ is an  $M \times 1$ column vector of the matrix product $\mathbf{AB}$  (in other words  $\mathbf{C}$), $\forall i=1, \dots, M$.  Each $\mathbf{c}_{i}$ can be viewed as a linear combination of the $N$ columns of $\mathbf{A}$ with coefficients that are the entries of the column $\mathbf{b}_{i}$ of $\mathbf{B}$, i.e., 
\begin{align}
\mathbf{c}_{i} = \sum_{j=1}^{N} b_{j, i} \mathbf{a}_{j}.
\label{combination}
\end{align}
Now, in order to show that  $\mathbf{C}$ is almost surely full rank, { it suffices} to show that any $r(\mathbf{A})$ columns of $\mathbf{C}$  are linearly independent. Without loss of generality, let us pick the first $r(\mathbf{A})$ column vectors of  $\mathbf{C}$ and check the linear independence between these vectors, i.e.,  
\begin{align} 
\sum_{i=1}^{r(\mathbf{A})} \alpha_{i} \mathbf{c}_{i} = \mathbf{0}.
\label{linear_indep}
\end{align}
We say that these column vectors are linearly independent if and only if $\alpha_{i} = 0, \forall i=1, \dots, r(\mathbf{A})$. 
Using (\ref{combination}), we can write (\ref{linear_indep}) as
\begin{align}
\sum_{i=1}^{r(\mathbf{A})} \alpha_{i} \sum_{j=1}^{N} b_{j, i} a_{j} = \mathbf{0}. 
\label{new_comb}
\end{align}
Let us pick basis vectors of matrix $\mathbf{A}$ as $\mathbf{p}_{1}, \mathbf{p}_{2}, \dots, \mathbf{p}_{r(\mathbf{A})}$. It is worth noting that  $\mathbf{A}$ in (\ref{matrix_A}) can be decomposed into two full rank matrices (using the full rank decomposition Theorem \cite{horn2012matrix}) as follows:
\begin{align}
\mathbf{A}_{M \times N} = \mathbf{P}_{M \times r(\mathbf{A})} \mathbf{S}_{r(\mathbf{A}) \times N}, 
\end{align}
where $\mathbf{P}$ contains the basis vectors, i.e.,
\begin{align}
 \mathbf{P} =  \left[ \begin{matrix} \mathbf{p}_{1} & \mathbf{p}_{2} & \dots & \mathbf{p}_{r(\mathbf{A})} \end{matrix}  \right] ,
  \end{align}
and  $\mathbf{S}$ contains the spanning coefficients.  In other words, each column vector $\mathbf{a}_{j}$ can be written as a linear combination of the basis vectors as follows:
\begin{align}
\mathbf{a}_{j} = \sum_{k=1}^{r(\mathbf{A})} s_{k,j} \mathbf{p}_{k}. \label{basis_combination}
\end{align}
Plugging (\ref{basis_combination}) in (\ref{new_comb}) and doing simple algebraic manipulations, we have 
\begin{align}
 \mathbf{0} &= \sum_{i=1}^{r(\mathbf{A})} \alpha_{i} \sum_{j=1}^{N} b_{j, i}  \bigg( \sum_{k=1}^{r(\mathbf{A})} s_{k,j} \mathbf{p}_{k} \bigg) = \sum_{i=1}^{r(\mathbf{A})}\sum_{j=1}^{N}  \alpha_{i}  b_{j, i}  \bigg( \sum_{k=1}^{r(\mathbf{A})} s_{k,j} \mathbf{p}_{k} \bigg), \\
&= \sum_{k=1}^{r(\mathbf{A})} \underbrace{\bigg(\sum_{i=1}^{r(\mathbf{A})} \sum_{j=1}^{N}  \alpha_{i}  b_{j, i}    s_{k,j}\bigg)}_{\lambda_{k}} \mathbf{p}_{k}. \label{equation_linear_basis}
\end{align}
Since the basis vectors are linear independent (i.e., $\lambda_{k} = 0, \forall k = 1, \dots, r(\mathbf{A})$), we have
\begin{align}
\lambda_{k} = \sum_{i=1}^{r(\mathbf{A})} \sum_{j=1}^{N}  \alpha_{i}  b_{j, i}    s_{k,j} = 0, \hspace{0.05in} \forall k =1, \dots, r(\mathbf{A}).
\end{align}
The previous equation can be written in a matrix form as follows: 
\begin{align}
\left[ \begin{matrix} \sum_{j=1}^{N} b_{j,1} s_{1,j} & \dots & \sum_{j=1}^{N} b_{j, r(\mathbf{A})} s_{1, j} \\ \vdots & \ddots & \vdots 
 \\ \sum_{j=1}^{N} b_{j,1} s_{r(\mathbf{A}), j} & \dots & \sum_{j=1}^{N}b_{j, r(A)} s_{r(\mathbf{A}), j}\end{matrix} \right]_{r(\mathbf{A}) \times r(\mathbf{A}) }  \begin{bmatrix}
\alpha_{1} \\ \alpha_{2}  \\ \vdots \\ \alpha_{r(\mathbf{A})} \end{bmatrix} = \begin{bmatrix}
0 \\ 0 \\ \vdots \\ 0 \end{bmatrix}, \label{matrix_equation}
\end{align}
which can be re-written as
\begin{align}
{\underbrace{\left[ \begin{matrix}  s_{1,1} & s_{1,2} &\dots & s_{1, N} \\ 
s_{2,1 } & s_{2,2} & \dots & s_{2, N} \\ \vdots & \vdots & \ddots & \vdots \\
s_{r(\mathbf{A}),1 } & s_{r(\mathbf{A}),2} & \dots & s_{r(\mathbf{A}),N }   
\end{matrix} \right]}_{\mathbf{S}}}  { \underbrace{\left[ \begin{matrix}  b_{1,1} & b_{1,2} &\dots & b_{1, r(\mathbf{A})} \\ 
b_{2,1 } & b_{2,2} & \dots & b_{2, r(\mathbf{A})} \\ \vdots & \vdots & \ddots & \vdots \\
b_{N,1 } & b_{N,2} & \dots & b_{N, r(\mathbf{A}) }   
\end{matrix} \right]}_{\hat{\mathbf{B}} }} \begin{bmatrix}
\alpha_{1} \\ \alpha_{2}  \\ \vdots \\ \alpha_{r(\mathbf{A})} \end{bmatrix} = \begin{bmatrix}
0 \\ 0 \\ \vdots \\ 0 \end{bmatrix}. \label{yala}
\end{align}
\noindent If  $\mathbf{S} \hat{\mathbf{B}}$ is full rank, this will imply that $\alpha_{i} = 0, \forall i = 1, 2, \dots, r(\mathbf{A})$, which in turn will imply that $\mathbf{C} = \mathbf{AB}$ is full rank.  Hence, our goal is to show that the matrix product $\mathbf{S} \hat{\mathbf{B}}$ is full rank, i.e., 
\begin{align}
\text{rank}(\mathbf{S} \hat{\mathbf{B}}) = r(\mathbf{A}), \hspace{0.05in} \text{almost surely}.
\end{align}
We have, 
\begin{align}
\mathbf{S} =  \left[ \begin{matrix} \mathbf{s}_{1} & \mathbf{s}_{2} & \dots & \mathbf{s}_{N} \end{matrix}  \right],  \hspace{0.1in} \hat{\mathbf{B}} &= \left[ \begin{matrix} \mathbf{b}_{1} & \mathbf{b}_{2} & \dots & \mathbf{b}_{r(\mathbf{A})} \end{matrix}  \right].
\end{align} 
Now, the matrix product can be written as 
\begin{align}
\mathbf{S} \hat{\mathbf{B}} =  \mathbf{D} =  \left[ \begin{matrix} \mathbf{d}_{1} & \mathbf{d}_{2} & \dots & \mathbf{d}_{r(\mathbf{A})} \end{matrix}  \right]. \label{nafea}
\end{align}
Then, we need to check the linear independence condition, 
\begin{align}
\sum_{i=1}^{r(\mathbf{A})} \beta_{i} \mathbf{d}_{i} = \mathbf{0},
\end{align}
where $\mathbf{d}_{i} = \sum_{j=1}^{N} b_{j, i} \mathbf{s}_{j}$. Now, we do simple algebraic arrangements as follows:
\begin{align}
\mathbf{0} &= \sum_{i=1}^{r(\mathbf{A})} \beta_{i}  \sum_{j=1}^{N} b_{j, i} \mathbf{s}_{j} = \sum_{i=1}^{r(\mathbf{A})}   \sum_{j=1}^{N} \beta_{i}  b_{j, i} \mathbf{s}_{j} =  \sum_{j=1}^{N} \underbrace{\bigg( \sum_{i=1}^{r(\mathbf{A})}  \beta_{i}  b_{j, i}\bigg)}_{m_{j}} \mathbf{s}_{j}.  \label{important_eqn}
\end{align}
The coefficients $\{m_{j}\}_{j=1}^{N}$ are functions of $\{\beta_{i}\}_{i=1}^{r(\mathbf{A})}$ and $b_{j, i},  j = 1, \dots, N$ and $i = 1, \dots, r(\mathbf{A})$ which can be written in a matrix form as follows: 
\begin{align}
\label{matrix_equation_2}
 \begin{bmatrix}
m_{1} \\ m_{2} \\ \vdots \\ m_{N} \end{bmatrix} = \left[ \begin{matrix}  b_{1,1} & b_{1,2} &\dots & b_{1, r(\mathbf{A})} \\ 
b_{2,1 } & b_{2,2} & \dots & b_{2, r(\mathbf{A})} \\ \vdots & \vdots & \ddots & \vdots \\
b_{N,1 } & b_{N,2} & \dots & b_{N, r(\mathbf{A}) }   
\end{matrix} \right]_{N \times r(\mathbf{A})} \begin{bmatrix}
\beta_{1} \\ \beta_{2}  \\ \vdots \\ \beta_{r(\mathbf{A})} \end{bmatrix} .
\end{align}
The $N$ columns of $\mathbf{S}$ are linearly dependent since the rank of $\mathbf{S}$ is $r(\mathbf{A})$, and each column is  of length $r(\mathbf{A})$, where $r(\mathbf{A}) < N$. Therefore, the matrix equation (\ref{matrix_equation_2}) has infinitely many solutions for $\{m_{j}\}_{j=1}^{N}$. Since the number of equations ($N$) is greater than the number of unknowns ($r(\mathbf{A})$), this has a solution for $\{\beta_{i}\}_{i=1}^{r(\mathbf{A})}$ if and only if the elements $b_{j,i}, j=1,2, \dots, N, i=1,2, \dots, r(\mathbf{A})$ have some structure, i.e., they are dependent. Since the entries of $\mathbf{B}$ are independently drawn from some continuous distribution, the probability that these entries being dependent is zero. Moreover, consider the set with inifinite cardinality, where each element in this set is a structured set $\mathbf{B}$ that causes the system of equations in  (\ref{matrix_equation_2}) to have a solution for $\{\beta_{i}\}_{i=1}^{r(\mathbf{A})}$, for some $\{m_{j}\}_{j=1}^{N}$. This set with infinite cardinality has a Lebesgue measure zero in the space $\mathds{F}^{N \times r(\mathbf{A})}$ (where $\mathds{F}$ is a field, i.e., $\mathds{R}$ or $\mathds{C}$) since this set is a subspace of $\mathds{F}^{N \times r(\mathbf{A})}$ with a dimension strictly less than $N \times r(\mathbf{A})$. Hence, we conclude that (\ref{nafea}) has no non-zero solution for $\{\beta_{i}\}_{i=1}^{r(\mathbf{A})}$  almost surely. Thus the matrix product $\mathbf{S}\hat{\mathbf{B}}$ is almost surely full rank and invertible. Then, from (\ref{yala}), it follows that,  the coefficients $\alpha_{i}, \forall i=1, \dots, r(\mathbf{A})$  are  zeros and consequently, 
\begin{align}
\text{rank}(\mathbf{AB}) = r(\mathbf{A}), \hspace{0.05in} \text{almost surely}.
\end{align}
This completes the proof of Lemma \ref{lemma_2}.

\section{Stochastic Encoding \& Equivocation Analysis} \label{appendix_stochastic}

Now we show the equivocation analysis. Our transmission works as follows:  We employ our transmission scheme over transmission block of length $B = RT + K(n+1)^{N}$. We apply stochastic encoding over $\tau$ transmission blocks  (i.e., we repeat the scheme over $\tau$ times). Fig. \ref{stochastic_encoding_block} gives an overview for our analysis.

 \begin{figure}[h]
  \centering
\includegraphics[scale=0.5]{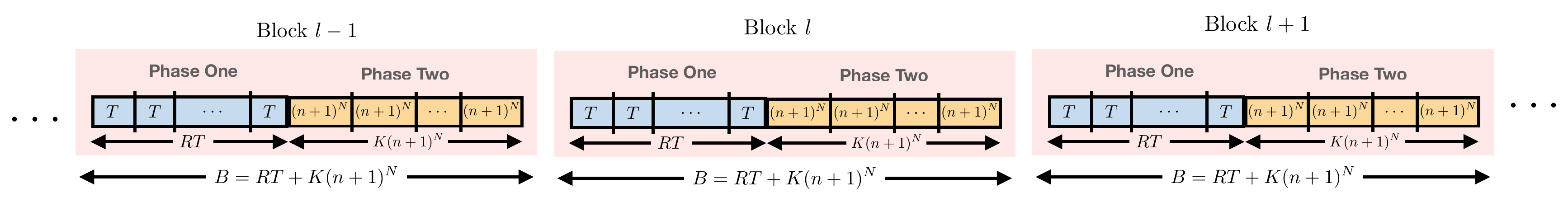}
\caption{Stochastic encoding over transmission blocks.}
\label{stochastic_encoding_block}
\end{figure}

To prove the secrecy for each message $W_{i}$. We will follow similar steps as in \cite{xie2015secure}.  We include the proof here for the sake of completeness. Throughout the analysis, non-boldface capital letters denote scalar random variables, and their values with non-boldface small letters. Also,  we denote $\tau$-length random variables with boldface capital letters, and their values with boldface small letters.

 Our goal is to show that when transmitting over many transmission blocks, we can drive the probability of error and the information leakage to zero as the number of transmission  blocks tends to infinity. We start the analysis by enhancing the eavesdroppers by conditioning on the unintended  information symbols  $(\text{i.e.},  S_{-i}^{K})$. Therefore, 
\begin{align}
I(S_{i}; Y_{j}) \leq  I (S_{i}; Y_{j} | S_{-i}^{K}) = I(S_{i}; Y_{j}, S_{-i}^{K}) \triangleq I(S_{i}; \tilde{Y}_{j}),
\end{align}
where $\tilde{Y}_{j} \triangleq (Y_{j}, S_{-i}^{K})$ is the output of an enhanced eavesdropper $j$ with respect to message $W_{i}$.   We will assume that $X$, $Y$ are qunatized versions of their original values using the discretization procedure as described in \cite{el2011network}, then we can use the discrete entropy in the equivocation analysis. It is worth noting that the analysis after discretization is equivalent to the original problem when the quantization step, $\Delta \rightarrow 0$, (see, Theorem 3.3 \cite{el2011network}). Hence, we can use the DMC achievability scheme as described in \cite{xie2015secure}.

      \textbf{Note:} The input of the channel $\textbf{X}$ of block $l$ is a multi-dimensional vector of size $B$, however, for notation convenience in the equivocation analysis we will treat $\textbf{X}$ as a scalar random variable and that will not change the analysis. Now, we will treat $\textbf{X}$ as the input of the channel $\tau$ transmission blocks.

  
Now, we want to prove secrecy for each message $W_{i}$, via the following equivocation inequality:  
\begin{align}
H (W_{i} | \mathbf{Y}_{j}, \mathbf{S}_{-i}^{K} ) \geq H(W_{i}) - \tau \epsilon^{(i)}, \hspace{0.05in} i=1,2, \dots, K, j\neq i  \label{fact}
\end{align}
for arbitrarily small $\epsilon^{(i)}$. 

\noindent Therefore, equation (\ref{fact}) implies the original secrecy constraints in  (3) and (4) from the following:
\begin{align}
H(W_{-j}^{K} | \mathbf{Y}_{j}) & \geq H(W_{-j}^{K} | \mathbf{Y}_{j}, W_{j}), \\
& \geq \sum_{i \neq j} H(W_{i} | \mathbf{Y}_{j}, \mathbf{S}_{-i}^{K}, W_{-i}^{K}), \\ 
& = \sum_{i \neq j} H(W_{i} | \mathbf{Y}_{j}, \mathbf{S}_{-i}^{K}), \\
& \overset{(a)} \geq \sum_{i \neq j} \left[ H(W_{i} - \tau \epsilon^{(i)})\right], \\
& = H(W_{-j}^{K}) - \tau \epsilon^{(-j)},
\end{align}
where $\epsilon^{(-j)} = \sum_{i \neq j} \epsilon^{i}$. Step (a) is due to the Markov chain $W_{-i}^{K} \rightarrow (\mathbf{Y}_{j}, \mathbf{S}_{-i}^{K}) \rightarrow W_{i}$. Similarly, 
\begin{align}
H(W^{K} | \mathbf{Z}) & \geq \sum_{i} H(W_{i} | \mathbf{Z}, W_{-i}^{K} ), \\
& \geq H (W_{i} | \mathbf{Z}, \mathbf{S}_{-i}^{K}), \\ 
& = \sum_{i} H(W_{i} | \mathbf{Z}, \mathbf{S}_{-i}^{K}), \\
& \geq \sum_{i} \left[ H(W_{i}) - \tau \epsilon^{(i)} \right], \\ 
& = H(W^{K}) - \tau \epsilon^{(z)}
\end{align}
where $\epsilon^{(z)} = \sum_{i} \epsilon^{i}$, is small for sufficiently large $\tau$.

\subsection*{Codebook Generation}

We consider the achievable secrecy rate against the \textit{strongest} adversary (i.e., $K-1$ receivers and the external eavesdropper) as in [R3]. For each transmitter $i$, we construct a compound wiretap code. We first choose the following rates for the secure and confusion messages of each transmitter $i$ as follows: 
\begin{align}
& R_{i} = I (X_{i}; Y_{i}) -   \max \left[  \max_{j} I(S_{i}; Y_{j} | S_{-i}^{K}), I(S_{i}; Z | S_{-i}^{K})  \right]  - \epsilon, \\
& R_{i}^{c} =  \max \left[  \max_{j} I(S_{i}; Y_{j} | S_{-i}^{K}), I(S_{i}; Z | S_{-i}^{K})  \right]  - \epsilon.
\end{align}
Each transmitter $i$ generates $2^{\tau (R_{i} + R_{i}^{c})}$ typical sequences $\mathbf{s}_{i}$ each with probability $\text{Pr}(\mathbf{s}_{i}) = \prod_{t=1}^{\tau} \text{Pr}(s_{it})$. For each transmitter $i$, we construct a codebook as follows:
\begin{align}
C_{i} \triangleq \{ \mathbf{s}_{i}(w_{i}, w_{i}^{c}): w_{i} \in \{1, \dots, 2^{\tau R_{i}}\},  w_{i}^{c} \in \{1, \dots, 2^{\tau R_{i}^{c}}\} \}.
\end{align} 
To transmit a message $w_{i}$, transmitter $i$ chooses an element $\mathbf{v}_{i}$ from the sub-codebook $C_{i}(w_{i})$ as follows: 
\begin{align}
C_{i}(w_{i}) \triangleq \{ \mathbf{s}_{i}(w_{i}, w_{i}^{c}):  w_{i}^{c} \in \{1, \dots, 2^{\tau R_{i}^{c}}\} \}
\end{align} 
and generate a channel input sequence with probability $\text{Pr}(x_{i} | s_{i})$. Since, 
\begin{align}
R_{i} + R_{i}^{c} < I (S_{i}; Y_{i})
\end{align}
therefore for sufficiently large $\tau$, the probability of error at receiver $i$ can be bounded by $\epsilon$, i.e., $\text{Pr}(e_{i})^{(\tau)} \leq \epsilon  $.






Now our goal is to lower bound $H(W_{i} | \mathbf{Y}_{j}, \mathbf{S}_{-i}^{K})$. Before we proceed the analysis, we write $H(W_{i} | \mathbf{Y}_{j}, \mathbf{S}_{-i}^{K})$ as follows:
\begin{align}
H(W_{i} | \mathbf{Y}_{j}, \mathbf{S}_{-i}^{K}) & = H(W_{i}, \mathbf{Y}_{j} | \mathbf{S}_{-i}^{K}) - H(\mathbf{Y}_{j} | \mathbf{S}_{-i}^{K}), \\ 
&=H(W_{i}, \mathbf{S}_{i}, \mathbf{Y}_{j} | \mathbf{S}_{-i}^{K}) - H(\mathbf{S}_{i} | W_{i}, \mathbf{Y}_{j}, \mathbf{S}_{-i}^{K}) - H(\mathbf{Y}_{j} | \mathbf{S}_{-i}^{K}), \\
& = H(W_{i}, \mathbf{S}_{i} | \mathbf{S}_{-i}^{K}) + H(\mathbf{Y}_{j} | W_{i}, \mathbf{S}_{i}, \mathbf{S}_{-i}^{K}) - H(\mathbf{S}_{i} | W_{i}, \mathbf{Y}_{j}, \mathbf{S}_{-i}^{K}) - H(\mathbf{Y}_{j} | \mathbf{S}_{-i}^{K}),\\  
& \overset{(a)} = H(W_{i}, \mathbf{S}_{i} | \mathbf{S}_{-i}^{K}) - H(\mathbf{S}_{i} | W_{i}, \mathbf{Y}_{j}, \mathbf{S}_{-i}^{K}) + H(\mathbf{Y}_{j} | \mathbf{S}_{i}, \mathbf{S}_{-i}^{K}) - H(\mathbf{Y}_{j} | \mathbf{S}_{-i}^{K}), \\ 
& \overset{(b)} = H({W}_{i}) - H(\mathbf{S}_{i} | W_{i}, \mathbf{Y}_{j}, \mathbf{S}_{-i}^{K}) + H(\mathbf{Y}_{j} | \mathbf{S}_{i}, \mathbf{S}_{-i}^{K}) - H(\mathbf{Y}_{j} | \mathbf{S}_{-i}^{K}).  \label{bounding}
\end{align}
In (a), the third term forms a Markov chain $W_{i} \rightarrow (\mathbf{S}_{i}, \mathbf{S}_{-i}^{K}) \rightarrow \mathbf{Y}_{j}$.  The first term in (a) is written as follows: 
\begin{align}
H(W_{i}, \mathbf{S}_{i} | \mathbf{S}_{-i}^{K}) &= H(W_{i}, \mathbf{S}_{i}), \\ 
& = H(\mathbf{S}_{i}) +  \underbrace{H(  W_{i}  | \mathbf{S}_{i} )}_{=0}.
\end{align}
Therefore, we have step (b). Note that,
\begin{align}
 H({\mathbf{S}}_{i}) = \tau (R_{i} + R_{i}^{c}). \label{sub_cardinality}
\end{align}

 Now we want to bound the second term in (b). Given the message $W_{i} = w_{i}$ and the received sequences $\mathbf{Y}_{j} = \mathbf{y}_{j}$ and the genie-aided sequences $\mathbf{S}_{-i}^{K} = \mathbf{s}_{-i}^{K}$, receiver $Y_{j}$ can decode the codeword $\mathbf{s}_{i}(w_{i}, w_{i}^{c})$ with arbitrarily small probability of error $\lambda(w_{i})^{(\tau)}$ as $\tau$ gets large.


Without loss of generality, assume that $\mathbf{s}_{i} (w_{i}, w_{1}^{c})$ is sent.  Error is defined as 
\begin{align}
E_{j} \triangleq \{ (\mathbf{s}_{i}(w_{i}, w_{j}^{c}), \mathbf{y}_{j} )  \in \mathcal{T}_{\epsilon}^{(\tau)} (P_{\mathbf{S}_{1}, \mathbf{Y}_{j} | \mathbf{S}_{-i}^{K}}) (\mathbf{s}_{-i}^{K}) \}
\end{align}
The probability of error $\lambda(w_{i})^{(\tau)} $ is bounded as follows:
\begin{align}
\lambda(w_{i})^{(\tau)} \leq \text{Pr} (E_{1}^{c}) + \sum_{j \neq 1} \text{Pr} (E_{j}),
\end{align}
where the probability here is conditioned on the event that $\mathbf{s}_{i}(w_{i}, w_{1}^{c})$ is sent. 
Note that, 
\begin{align}
\text{Pr} (E_{1}^{c}) \leq \epsilon_{1}
\end{align}
for sufficiently large $\tau$, and
\begin{align}
\text{Pr} (E_{j})  & \leq 2^{\tau H(S_{i}, Y_{j} | S_{-i}^{K}) - \tau H(S_{i}) - \tau H(Y_{j} | S_{-i}^{K}) - \tau \epsilon_{2}}, \\
& = 2^{- \tau I (S_{1}; Y_{j} | S_{-i}^{K} ) - \tau \epsilon_{2}}
\end{align}
Hence,
\begin{align}
\lambda(w_{i})^{(\tau)}  & \leq \epsilon_{1} + 2^{\tau R_{i}^{c}} 2^{- \tau I (S_{1}; Y_{j} | S_{-i}^{K}) - \tau \epsilon_{2}}, \\ 
& \leq  \epsilon_{3}.
\end{align}
By using Fano's inequality,  we have the following:
\begin{align}
H(\mathbf{S}_{i} | W_{i}, \mathbf{Y}_{j}, \mathbf{S}_{-i}^{K} )  & \leq (1 + \lambda(w_{i})^{(\tau)}  \log 2^{\tau (R_{i} + R_{i}^{c})} ), \\ 
& \leq \tau \epsilon_{4} \label{sub_fano}
\end{align}

Now we bound the third term in (\ref{bounding}) as follows:
\begin{align}
& H (\mathbf{Y}_{j} | \mathbf{S}_{i}, \mathbf{S}_{-i}^{K})  = \sum_{\mathbf{s}_{i}, \mathbf{s}_{-i}^{K}} \text{Pr} (\mathbf{S}_{i} = \mathbf{s}_{i} )  \text{Pr} (\mathbf{S}_{-i}^{K} = \mathbf{s}_{-i}^{K}) H(\mathbf{Y}_{j} | \mathbf{S} _{i} = \mathbf{s}_{i}, \mathbf{S}_{-i}^{K} = \mathbf{s}_{-i}^{K}), \\ 
& \geq  \sum_{\mathbf{s}_{i}, \mathbf{s}_{-i}^{K} \in \mathcal{T}_{\epsilon}^{(\tau) } (P(s_{i}, s_{-i}^{K} ) )} \left[   \text{Pr} (\mathbf{S}_{i} = \mathbf{s}_{i} )  \text{Pr} (\mathbf{S}_{-i}^{K} = \mathbf{s}_{-i}^{K}) H(\mathbf{Y}_{j} | \mathbf{S} _{i} = \mathbf{s}_{i}, \mathbf{S}_{-i}^{K} = \mathbf{s}_{-i}^{K}) \right] , \\
& \geq  \sum_{\mathbf{s}_{i}, \mathbf{s}_{-i}^{K} \in \mathcal{T}_{\epsilon}^{(\tau) } (P(s_{i}, s_{-i}^{K} ) )}  \text{Pr} (\mathbf{S}_{i} = \mathbf{s}_{i} )  \text{Pr} (\mathbf{S}_{-i}^{K} = \mathbf{s}_{-i}^{K}) \nonumber \\
& \hspace{0.3in} \times   \sum_{(a, b) \in \mathcal{S}_{i} \times \mathcal{S}_{-i}^{K}} N(a, b | \mathbf{s}_{i}, \mathbf{s}_{-i}^{K})   \sum_{y_{j} \in \mathcal{Y}_{j}}  - \text{Pr} (y_{j} | a, b) \log(\text{Pr} (y_{j} | a, b)) , \\
& \geq  \sum_{\mathbf{s}_{i}, \mathbf{s}_{-i}^{K} \in \mathcal{T}_{\epsilon}^{(\tau) } (P(s_{i}, s_{-i}^{K} ) )}  \text{Pr} (\mathbf{S}_{i} = \mathbf{s}_{i} )  \text{Pr} (\mathbf{S}_{-i}^{K} = \mathbf{s}_{-i}^{K})  \nonumber \\ 
&  \hspace{0.3in} \times  \sum_{(a, b) \in \mathcal{S}_{i} \times \mathcal{S}_{-i}^{K}} \tau \left( \text{Pr} (S_{i} = a, S_{-i}^{K} = b) - \epsilon_{5}  \right)  \nonumber \\ 
  & \hspace{0.3in} \times  \sum_{y_{j} \in \mathcal{Y}_{j}}  - \text{Pr} (y_{j} | a, b) \log(\text{Pr} (y_{j} | a, b)), \\
& \geq  \sum_{\mathbf{s}_{i}, \mathbf{s}_{-i}^{K} \in \mathcal{T}_{\epsilon}^{(\tau) } (P(s_{i}, s_{-i}^{K} ) )}  \tau \left[ \text{Pr} (\mathbf{S}_{i} = \mathbf{s}_{i} )  \text{Pr} (\mathbf{S}_{-i}^{K} = \mathbf{s}_{-i}^{K}) H(Y_{j} | s_{i}, s_{-i}^{K} - \epsilon_{6} ) \right], \\
& \geq (1-\epsilon_{7}) \tau H(Y_{j} | s_{i}, s_{-i}^{K}) - \tau \epsilon_{8}, \\ 
& \geq \tau H(Y_{j} | s_{i}, s_{-i}^{K}) - \tau \epsilon_{9}.
\end{align}

Now we upper bound the fourth term in (\ref{bounding}). First, let us define 
\begin{align}
\hat{\mathbf{Y}}_{j}= \begin{cases}
 \mathbf{Y}_{j}, & \text{if } (\mathbf{s}_{-i}^{K}, \mathbf{y}_{j}) \in \mathcal{T}_{\epsilon}^{(\tau)} (P_{S_{-i}^{K}, Y_{j} }),    \\
\text{arbitrary}, & \text{otherwise. } 
\end{cases}
\end{align}
Then, we have the following:
\begin{align}
H(\mathbf{Y}_{j} | \mathbf{s}_{-i}^{K}) & = \sum_{\mathbf{s}_{-i}^{K}} \text{Pr} (\mathbf{S}_{-i}^{K} = \mathbf{s}_{-i}^{K}  ) H(\mathbf{Y}_{j} | \mathbf{S}_{-i}^{K} = \mathbf{s}_{-i}^{K} ), 
\end{align}
\begin{align}   
& \leq \sum_{\mathbf{s}_{-i}^{K}} \text{Pr} (\mathbf{S}_{-i}^{K} = \mathbf{s}_{-i}^{K}  ) H( \hat{\mathbf{Y}}_{j}, \mathbf{Y}_{j} | \mathbf{S}_{-i}^{K} = \mathbf{s}_{-i}^{K} ), \\  
& = \sum_{\mathbf{s}_{-i}^{K}} \text{Pr} (\mathbf{S}_{-i}^{K} = \mathbf{s}_{-i}^{K}  ) \left[ H( \hat{\mathbf{Y}}_{j}  | \mathbf{S}_{-i}^{K} = \mathbf{s}_{-i}^{K} ) + H(\mathbf{Y}_{j} |   \hat{\mathbf{Y}}_{j},  \mathbf{S}_{-i}^{K} = \mathbf{s}_{-i}^{K} ) \right]  , \\  
& \leq \tau H({Y}_{j} | {s}_{-i}^{K} ) + \tau \epsilon_{1} + \sum_{\mathbf{s}_{-i}^{K}} \text{Pr} (\mathbf{S}_{-i}^{K} = \mathbf{s}_{-i}^{K}  )   H(\mathbf{Y}_{j} |   \hat{\mathbf{Y}}_{j},  \mathbf{S}_{-i}^{K} = \mathbf{s}_{-i}^{K} ).
\end{align}
Combining Fano's inequality and the fact that 
\begin{align}
\text{Pr} (\mathbf{Y}_{j} \neq \hat {\mathbf{Y}}_{j} ) \leq \text{Pr} \left[ (\mathbf{s}_{-i}^{K}, \mathbf{Y}_{j})  \notin \mathcal{T}_{\epsilon}^{(\tau)} (P_{{S}_{-i}^{K}, Y_{j}} )  \right] 
\end{align}
is arbitrarily small for sufficiently large $\tau$, then 
\begin{align}
H(\mathbf{Y}_{j} | \mathbf{S}_{-i}^{K}) \leq \tau H({Y}_{j} | {S}_{-i}^{K}) + \tau \epsilon_{1} + \tau \epsilon_{2} \label{sub_final}
\end{align}
Substituting (\ref{sub_cardinality}), (\ref{sub_fano}) and  (\ref{sub_final}) into (\ref{bounding}), we conclude that
\begin{align}
H(W_{i} | \mathbf{Y}_{j}, \mathbf{S}_{-i}^{K} ) \geq H(W_{i}) - \tau \epsilon^{(i)}
\end{align}
where $\epsilon^{(i)}$ is small for sufficiently large $\tau$, which completes the proof.

\bibliographystyle{IEEEtran}
\bibliography{myreferences}

\end{document}